\documentclass[twocolumn,prb]{revtex4}
\usepackage{amsfonts}
\usepackage[T1]{fontenc}
\usepackage{amsmath,amsbsy,amssymb,graphicx}
\usepackage{times}

\let\mathbf=\boldsymbol
\input{tcilatex}

\begin{document}

\title{Non-Abelian braiding of Majorana-like edge states and\\
topological quantum computations in electric circuits}
\author{Motohiko Ezawa}
\affiliation{Department of Applied Physics, University of Tokyo, Hongo 7-3-1, 113-8656,
Japan}

\begin{abstract}
Majorana fermions subject to the non-Abelian braid group are believed to be
the basic ingredients of future topological quantum computations. In this
work, we propose to simulate Majorana fermions of the Kitaev model in
electric circuits based on the observation that the circuit Laplacian can be
made identical to the Hamiltonian. A set of AC voltages along the chain
plays a role of the wave function. We generate an arbitrary number of
topological segments in a Kitaev chain. A pair of topological edge states
emerge at the edges of a topological segment. Its wave function is
observable by the position and the phase of a peak in impedance measurement.
It is possible to braid any pair of neighboring edge states with the aid of
T-junction geometry. By calculating the Berry phase acquired by their
eigenfunctions, the braiding is shown to generate one-qubit and two-qubit
unitary operations. We explicitly construct Clifford quantum gates based on
them. We also present an operator formalism by regarding a topological edge
state as a topological soliton intertwining the trivial segment and the
topological segment. Our analysis shows that the electric-circuit approach
can simulate the Majorana-fermion approach to topological quantum
computations.
\end{abstract}

\maketitle

\section{Introduction}

The braiding relation plays a key role in future topological quantum
computations\cite{Moore,Das,Kitaev,TQC,SternA,Stern,NPJ,Bonder,Cheng,Pahomi}%
. Majorana-fermion edge states emerging in topological superconductors are
the best candidate\cite%
{AliceaBraid,Qi,Alicea,Lei,Been,Stan,Elli,Sato,Les,AA,DJ,Ivanov,Halperin,Sni}%
. Examples are the $p_{x}+ip_{y}$ topological superconductors\cite{Ivanov}
and the Kitaev $p$-wave topological superconductors\cite{AliceaBraid,DJ}.
However, an experimental realization of braiding of Majorana fermions still
remains challenging.

The study on topological phases started in condensed-matter physics\cite%
{Hasan,Qi}, but now expanded to other systems including acoustic\cite%
{TopoAco,Berto,Sss,He}, photoic\cite%
{KhaniPhoto,Hafe2,Hafezi,WuHu,TopoPhoto,Ozawa}, mechanical\cite{Nash,Huber}
and electric-circuit\cite{TECNature,ComPhys} systems. In particular, almost
all topological phases are known to be materialized in electric circuits
because the circuit Laplacian has the same expression as the Hamiltonian
when the circuit is appropriately designed\cite{TECNature,ComPhys}, where
the admittance corresponds to the energy. Very recently, we have generated
Majorana-like corner states akin to those in topological superconductors and
shown that the braiding of these Majorana-like corner states is possible in
electric circuits\cite{EzawaMajo}. Indeed, we have derived the relation $%
\sigma ^{2}=-1$, where $\sigma $ denotes a single exchange of two
topological corner states. It indicates that a topological corner state is
an Ising anyon. Note that the relation $\sigma ^{2}=1$ holds both for bosons
and fermions. However, the single exchange $\sigma $ is impossible in this
model. This is because the braiding is controlled by an applied field and
the direction of the field becomes opposite after a single braid. Another
problem is that it is not clear how to braid more than two topological
corner states.

\begin{figure}[t]
\centerline{\includegraphics[width=0.48\textwidth]{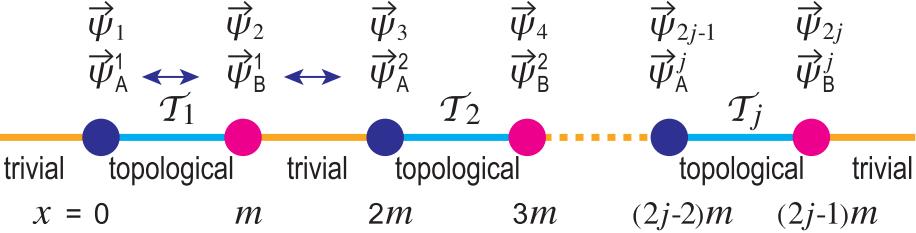}}
\caption{We create a number of topological segments on a single Kitaev chain
by controlling the chemical potential locally. A pair of topological
zero-energy states emerge at the edges of the $j$-th topological segment $%
\mathcal{T}_{j}$ (cyan). Their wave functions are denoted by $\vec{\protect%
\psi}_{A}^{j}$ and $\vec{\protect\psi}_{B}^{j}$. We may relabel them as $(%
\vec{\protect\psi}_{A}^{1},\vec{\protect\psi}_{B}^{1},\vec{\protect\psi}%
_{A}^{2},\vec{\protect\psi}_{B}^{2},\cdots )\rightarrow (\vec{\protect\psi}%
_{1},\vec{\protect\psi}_{2},\vec{\protect\psi}_{3},\vec{\protect\psi}%
_{4},\cdots )$. We investigate two basic braids: (i) We braid two edge
states $\vec{\protect\psi}_{A}^{1}$ and $\vec{\protect\psi}_{B}^{1}$ across
a topological segment, leading to a unitary operation in one qubit. (ii) We
braid two edge states $\vec{\protect\psi}_{B}^{1}$ and $\vec{\protect\psi}%
_{A}^{2}$ across a trivial segment, leading to a unitary operation in two
qubits. For definiteness, it is assumed that each topological segment
contains $m+1$ sites while each trivial segment contains $m-1$ sites. The
site index is expressed as $x=0,1,2,\cdots $. Edge states emerge at $%
x=0,m,2m,\cdots $. }
\label{FigBasicBraid}
\end{figure}

In this paper, employing the electric-circuit realization of the Kitaev model%
\cite{EzawaMajo}, we generate $N$ topological segments together with $N$
pairs of topological edge states in a Kitaev chain made of electric circuit.
We describe them by $N$ pairs of wave functions $(\vec{\psi}_{A}^{j},\vec{%
\psi}_{B}^{j})$, $j=1,2,\cdots ,N$, as in Fig.\ref{FigBasicBraid}. All pairs 
$(\vec{\psi}_{A}^{j},\vec{\psi}_{B}^{j})$ are orthogonal one to another,
since all topological segments are independent from one to another, thus
yielding $2^{N}$-fold degeneracy of the topological edge states.

The position and the phase of one edge state are observable by those of a
peak in impedance measurement. It is possible to carry out the braiding of
edge states with the use of T-junctions\cite{AliceaBraid,AA}. Furthermore,
we demonstrate that the braiding of two edge states across a topological
(trivial) segment generates one-qubit (two-qubit) unitary operation.

It is interesting to develop our scheme to pursue to what extent the
electric-circuit formalism can simulate the standard Majorana-operator
formalism. For this purpose, we remark that it is possible to regard a
topological edge state as a topological soliton because it intertwines the
topological and the trivial segments. Let us call it an edge soliton when we
focus on the aspect of soliton. We then define an operator that creates an
edge soliton. Remarkably, it behaves as a Majorana-fermion operator in the
electric-circuit formalism.

The paper is composed as follows. In Sec.~II, we focus on a single
topological segment together with a pair of edge solitons. A pair of wave
functions $\vec{\psi}_{A}$ and $\vec{\psi}_{B}$ are analytically constructed
to describe it. One-qubit states are described by superpositions of $\vec{%
\psi}_{A}$ and $\vec{\psi}_{B}$.

In Sec.~III, we review the electric-circuit realization of the Kitaev model.
A circuit consists of two main channels, i.e., the capacitor channel and the
inductor channel, corresponding to the electron band and the hole band in
the Kitaev model. It is shown that an edge soliton is observable by an
impedance peak both in these two channels, whose phase agrees precisely with
that of the wave function $\vec{\psi}_{A}$ or $\vec{\psi}_{B}$. We also
discuss how to register and observe the qubit information in the electric
circuit.

In Sec.~IV, we investigate the braiding of edge states. First, we analyze
how the eigenfunction evolves when some system parameters are locally
controlled. In particular, the Berry phase develops when the
"superconducting-phase" parameter is externally controlled. Then, using
these results, we investigate the braiding of two edges across a topological
segment and also across a trivial segment. Their effect is represented as
one-qubit and two-qubit unitary operators, respectively. We explicitly
construct Clifford quantum gates based on them.

In Sec.~V and VI, we introduce a creation operator $\gamma _{j}$ of an edge
soliton described by the wave function $\vec{\psi}_{j}$. It is argued that $%
\gamma _{j}$ is a Majorana-fermion operator by investigating the exchange
statistics of two edge solitons. All results are in consistent with those
derived by the analysis of the Berry phase.

In Sec.~VII, we present explicit formulas for the electric-circuit
realization of the Kitaev model by deriving the circuit Laplacian to be
identified with the Kitaev Hamiltonian. We also present an electric circuit
for a T-junction. Sec.~VIII is devoted to discussions.

\section{Kitaev model}

The Bogoliubov-de Gennes Hamiltonian is written as 
\begin{equation}
\hat{H}(\mathbf{k})=\Psi^{\dag }(\mathbf{k})H(\mathbf{k})\Psi (\mathbf{k}) ,
\label{2ndBdG}
\end{equation}%
with the Nambu operator 
\begin{equation}
\Psi (\mathbf{k})=\left\{ c(\mathbf{k}),c^{\dagger }(\mathbf{k})\right\} .
\label{NambuOpe}
\end{equation}%
It is customary to refer to $H(\mathbf{k})$ also as the Hamiltonian. The
Hamiltonian $H(k)$ may be regarded as a classical Hamiltonian, whereas $\hat{%
H}(k)$ is a second-quantized Hamiltonian. Topological properties of the
system are determined by the property of the classical Hamiltonian $H(k)$.

The Kitaev $p$-wave topological superconductor model is the fundamental
one-dimensional model hosting Majorana edge states\cite%
{Read,KitaevP,Kitaev,Alicea}. It is a two-band model whose Hamiltonian is%
\begin{equation}
H_{\text{K}}(k)=\frac{1}{2}\left( 
\begin{array}{cc}
\varepsilon _{k} & i\Delta e^{-i\phi }\sin k \\ 
-i\Delta e^{i\phi }\sin k & -\varepsilon _{k}%
\end{array}%
\right) ,  \label{BdG}
\end{equation}%
with%
\begin{equation}
\varepsilon _{k}=-t\cos k-\mu ,
\end{equation}
where $t$, $\mu $, $\phi $ and $\Delta $ represent the hopping amplitude,
the chemical potential, the superconducting phase and gap parameters,
respectively. It is well known that the system is topological for $%
\left\vert \mu \right\vert <\left\vert 2t\right\vert $ and trivial for $%
\left\vert \mu \right\vert >\left\vert 2t\right\vert $ irrespective of $%
\Delta $ provided $\Delta \neq 0$. A pair of topological zero-energy states
emerges at the edges of a topological phase according to the bulk-edge
correspondence. They are protected by particle-hole symmetry (PHS).

We consider a chain realizing the Kitaev model. A chain need not be
straight; it can bend or even branch off. We control the system parameters
locally so as to generate several topological segments with $\left\vert \mu
\right\vert <\left\vert 2t\right\vert $ sandwiched by trivial segments with $%
\left\vert \mu \right\vert >\left\vert 2t\right\vert $ as in Fig.\ref%
{FigBasicBraid}.

It is convenient to choose the parameters such that%
\begin{equation}
\Delta =t,\text{\qquad }\mu =0  \label{ParamTopo}
\end{equation}%
to generate a topological segment, and%
\begin{equation}
\Delta =t,\text{\qquad }\mu =4t  \label{ParamTrivial}
\end{equation}%
to generate a trivial segment. Then, we obtain analytical solutions
describing the zero-energy edge states as in Eqs.(\ref{GroundA}).

The Kitaev model (\ref{BdG}) is reduced to%
\begin{equation}
H_{\text{K}}^{y}(k)=\frac{1}{2}\varepsilon _{k}\sigma _{z}-\frac{1}{2}\Delta
\sigma _{y}\sin k  \label{Hy}
\end{equation}%
for $\phi =0$, and%
\begin{equation}
H_{\text{K}}^{x}(k)=\frac{1}{2}\varepsilon _{k}\sigma _{z}+\frac{1}{2}\Delta
\sigma _{x}\sin k.  \label{Hx}
\end{equation}%
for $\phi =\pi /2$. In the present work, we use the $H_{\text{K}}^{y}$ model
(\ref{Hy}), where $\phi =0$. However, this phase degree of freedom plays a
key role when we braid two edge states. The system parameters $t$, $\mu $, $%
\phi $ and $\Delta $ are locally controllable parameters in the
corresponding circuit Laplacian.

\subsection{Zero-energy solutions}

To obtain analytic solutions of the zero-energy states, we make a unitary
transformation of the Kitaev model (\ref{BdG}) as $H_{\text{K}}^{\prime
}\left( k\right) =U_{\text{K}}H_{\text{K}}\left( k\right) U_{\text{K}}^{-1}$
with 
\begin{equation}
U_{\text{K}}=\frac{1}{\sqrt{2}}\left( 
\begin{array}{cc}
-ie^{i\phi /2} & ie^{-i\phi /2} \\ 
e^{i\phi /2} & e^{-i\phi /2}%
\end{array}%
\right) ,  \label{UK}
\end{equation}%
and obtain%
\begin{equation}
H_{\text{K}}^{\prime }\left( k\right) =\frac{1}{2}\left( 
\begin{array}{cc}
0 & -i\varepsilon _{k}+\Delta \sin k \\ 
i\varepsilon _{k}+\Delta \sin k & 0%
\end{array}%
\right) .  \label{HK2}
\end{equation}%
With the choice (\ref{ParamTopo}) of the parameters, it is simplified as%
\begin{equation}
H_{\text{K}}^{\prime }\left( k\right) =\frac{1}{2}\left( 
\begin{array}{cc}
0 & ite^{-ik} \\ 
-ite^{ik} & 0%
\end{array}%
\right) .
\end{equation}%
When one topological segment contains $m+1$ sites, the zero-energy solutions
of this model are explicitly given in the coordinate space by%
\begin{equation}
\vec{\psi}_{A}^{\prime }=\left( 1,0,\cdots ,0\right) ,\qquad \vec{\psi}%
_{B}^{\prime }=\left( 0,\cdots ,0,1\right) ,
\end{equation}%
which are $2(m+1)$ component vectors.

By making the inverse unitary transformation, we obtain the zero-energy
solutions in the original Kitaev Hamiltonian (\ref{BdG}) as%
\begin{align}
\vec{\psi}_{A}& =U_{\text{K}}^{-1}\vec{\psi}_{A}^{\prime }  \notag \\
& =\frac{1}{\sqrt{2}}(ie^{-i\phi /2},0,\cdots ,0;-ie^{i\phi /2},0,\cdots ,0),
\notag \\
\vec{\psi}_{B}& =U_{\text{K}}^{-1}\vec{\psi}_{B}^{\prime }  \notag \\
& =\frac{1}{\sqrt{2}}(0,\cdots ,0,e^{-i\phi /2};0,\cdots ,0,e^{i\phi /2}),
\label{GroundA}
\end{align}%
where $\phi $ is the superconducting phase in the Hamiltonian (\ref{BdG}).
We refer to the first (last) $(m+1)$-components as the electron (hole)
sector in accord with the Nambu operator (\ref{NambuOpe}). It is seen that $%
\vec{\psi}_{A}$ and $\vec{\psi}_{B}$ are perfectly localized at the left
edge and the right edge, respectively. They agree with the wave functions of
the Majorana edge states in the Kitaev $p$-wave topological superconductor
model.

\subsection{One-qubit state}

\label{SecOneQubit}

We analyze a Kitaev chain containing one topological segment, where there
are two topological edge states $\vec{\psi}_{A}$ and $\vec{\psi}_{B}$. Any
linear combination of $\vec{\psi}_{A}$ and $\vec{\psi}_{B}$ is degenerate at
zero energy. We construct a set of orthogonal states $\vec{\psi}_{|0\rangle
} $ and $\vec{\psi}_{|1\rangle }$ as%
\begin{equation}
\left( 
\begin{array}{c}
\vec{\psi}_{|0\rangle } \\ 
\vec{\psi}_{|1\rangle }%
\end{array}%
\right) =U_{\text{basis}}\left( 
\begin{array}{c}
\vec{\psi}_{A} \\ 
\vec{\psi}_{B}%
\end{array}%
\right) ,  \label{EqD}
\end{equation}%
with%
\begin{equation}
U_{\text{basis}}=\frac{1}{\sqrt{2}}\left( 
\begin{array}{cc}
1 & i \\ 
1 & -i%
\end{array}%
\right) ,  \label{UAB}
\end{equation}%
which read%
\begin{align}
\vec{\psi}_{|n\rangle }& =\frac{i}{2}\{e^{-i\phi /2},\cdots ,0,\left(
-1\right) ^{n}e^{-i\phi /2};  \notag \\
& \qquad \qquad -e^{i\phi /2},\cdots ,0,\left( -1\right) ^{n}e^{i\phi /2}\},
\end{align}%
where $n=0,1$. As far as the electron sector concerns, the wave function $%
\vec{\psi}_{|0\rangle }$ is symmetric with respect to the change of the
components at $x=0$ and $x=m$, while $\vec{\psi}_{|1\rangle }$ is
antisymmetric. We later show that $\vec{\psi}_{|0\rangle }$ and $\vec{\psi}%
_{|1\rangle }$ are the wave functions describing one-qubit states $|0\rangle 
$ and $|1\rangle $, respectively: See Eq.(\ref{f-state}).

In application of the Kitaev model for quantum computation we start from and
end at the system with the "superconducting phase" $\phi =0$, where%
\begin{equation}
\vec{\psi}_{|n\rangle }=\frac{i}{2}(1,0,\cdots ,0,\left( -1\right)
^{n};-1,0,\cdots ,0,\left( -1\right) ^{n}).  \label{EqE}
\end{equation}%
We make the use of the phase degrees of freedom only when we perform
braiding of edge states.

\subsection{Multi-qubit state}

We proceed to consider a Kitaev chain containing $N$ topological segments,
where the $j$-th topological segment produces two topological edge states
described by the wave functions $\vec{\psi}_{A}^{j}$ and $\vec{\psi}_{B}^{j}$
as in Fig.\ref{FigBasicBraid}. We may construct a set of wave functions $%
\vec{\psi}_{|0\rangle }^{j}$ and $\vec{\psi}_{|1\rangle }^{j}$ as in Eq.(\ref%
{EqD}), describing one-qubit states $|0\rangle _{j}$ and $|1\rangle _{j}$
for the $j$-th topological segment. They are orthogonal and degenerate at
zero energy, forming a two-dimensional Hilbert space $\mathcal{T}_{j}$. When
there are $N$ topological segments, the total Hilbert space is the direct
product of $N$ Hilbert spaces, that is $\otimes _{j}\mathcal{T}_{j}$. The
many-body ground states are given by the direct product,%
\begin{equation}
\left\vert n_{1}n_{2}\cdots n_{N}\right\rangle =\left\vert
n_{1}\right\rangle _{1}\otimes \left\vert n_{2}\right\rangle _{2}\otimes
\cdots \otimes \left\vert n_{N}\right\rangle _{N}  \label{Qubit}
\end{equation}%
with $n_{j}=0,1$, where the index $j$ denotes the $j$-th topological
segment. The ground-state degeneracy is $2^{N}$ as in the case of
topological superconductors, although we have derived it solely based on the
classical Hamiltonian.

\begin{figure}[t]
\centerline{\includegraphics[width=0.28\textwidth]{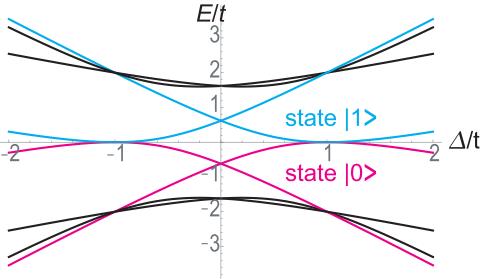}}
\caption{Energy $E/t$ as a function of $\Delta /t$ for a finite chain with
length 4. The two zero-energy states split and acquire nonzero energy for $%
t\neq \Delta $.}
\label{FigSplit}
\end{figure}

\subsection{Initialization}

\label{SecIni}

It is standard to start with the pure state $\left\vert 00\cdots
0\right\rangle $ to carry out quantum computation. Such a pure state can be
prepared as follows. We first tune the "superconducting" gap $|\Delta |$
slightly larger than the hopping amplitude $|t|$, $|\Delta /t|>1$, in a
topological segment. Although the two-fold degeneracy of the state $%
\left\vert 0\right\rangle $ and the state $\left\vert 1\right\rangle $ is
intact in an infinitely long system due to the PHS, it is broken in a finite
system because there is a mixing between the two edge states. The energy of
the state $\left\vert 0\right\rangle $ becomes lower than that of the state $%
\left\vert 1\right\rangle $, as numerically shown in Fig.\ref{FigSplit} for
a finite chain with length 4. Thus, we can choose the state $\left\vert
0\right\rangle $. By doing this setup for all topological phases, we can
construct the pure state $\left\vert 00\cdots 0\right\rangle $.

\begin{figure}[t]
\centerline{\includegraphics[width=0.48\textwidth]{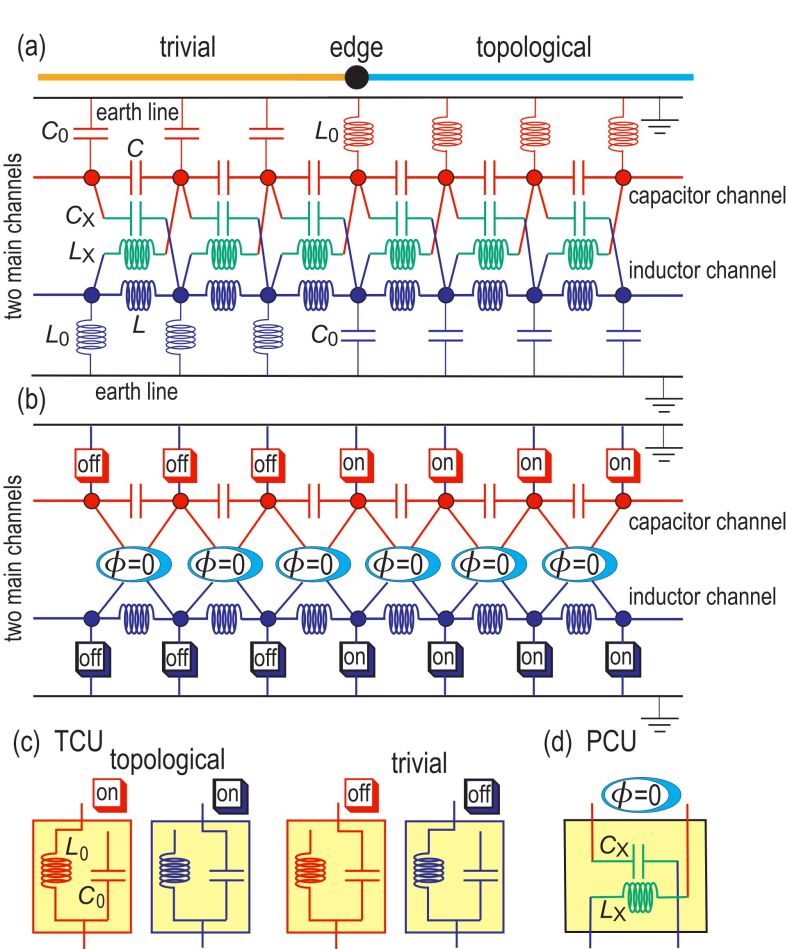}}
\caption{(a) A topological edge state emerges at the boundary between a
trivial and a topological segment. This Kitaev chain is simulated by a set
of two channels containing capacitors \textit{C} (red) and inductors \textit{%
L} (blue), corresponding to the electron band and the hole band,
respectively. Each site is connected to the ground via a capacitor ($C_{0}$)
or an inductor ($L_{0}$) to realize trivial and topological segments as
shown in this figure. In the model with $\protect\phi =0$, the two main
channels are crosslinked by a circuit composed of capacitors ($C_{X}$) and
inductors ($L_{X}$). (b) The same Kitaev chain is illustrated in terms of
the TCU (topology-control unit) and the PCU (phase-control unit) with $%
\protect\phi =0$. (c) Illustration of TCU. (d) Illustration of PCU at $%
\protect\phi =0$. }
\label{FigCircuitTT}
\end{figure}

\section{Electric-circuit realization}

\label{SecEC}

Electric circuits can simulate various topological systems\cite%
{TECNature,ComPhys,Hel,Lu,Research,Zhao,EzawaTEC,Garcia,Hofmann,EzawaLCR,Haenel,EzawaSkin,XiaoXiao}%
. We review how to realize the Kitaev model by an electric circuit\cite%
{EzawaMajo}. We use two main channels (red and blue) to represent a two-band
model as in Fig.\ref{FigCircuitTT}(a): One channel (i.e., capacitor channel)
consists of capacitors $C$ in series, implementing the electron band, while
the other channel (i.e., inductor channel) consists of inductors $L$ in
series, implementing the hole band. The hopping parameters are represented
by capacitors $C$ and inductors $L$, whose contribution to the circuit
Laplacian is $i\omega C$ and $1/i\omega L$: See Eq.(\ref{Kirch}). It is
understood that the hopping parameters are\ opposite between the electron
band and the hole band. We then introduce a pairing interaction between
them, by crosslinking the two main channels with the use of capacitors $%
C_{X} $ and inductors $L_{X}$, as shown in green in Fig.\ref{FigCircuitTT}%
(a). Each site in the main channel is connected to the ground via an
inductor $L_{0}$ or a capacitor $C_{0}$ as shown in Fig.\ref{FigCircuitTT}%
(a). The use of $C_{0}$ in the capacitor channel and $L_{0}$ in the inductor
channel makes the segment trivial, while the use of $L_{0}$ in the capacitor
channel and $C_{0}$ in the inductor channel makes the segment topological.

It is convenient to simplify Fig.\ref{FigCircuitTT}(a) down to Fig.\ref%
{FigCircuitTT}(b), where we have introduced subcircuits called the
topology-control unit (TCU) and the phase-control unit (PCU) explained in
Figs.\ref{FigCircuitTT}(c) and (d). When all TCUs are set on (off) in a
segment, the segment is in the topological (trivial) phase. A topological
edge state emerges at an edge site of a topological segment. By switching on
a TCU attached to the adjacent site, a topological edge state is shifted to
the adjacent site. Namely, we can move a topological edge state freely along
a Kitaev chain. In a braiding process of two edge states, as we explain in
Sec.\ref{PCU}, it is necessary to control the "superconducting phase" $\phi $
present in the Kitaev model (\ref{BdG}), which is controlled with the use of
PCUs.

\subsection{Circuit Laplacian}

\label{SecCL}

Electric circuits are characterized by the Kirchhoff current law\cite%
{TECNature,ComPhys,Hel}, 
\begin{align}
\frac{d}{dt}I_{a} &=\sum_{b}C_{ab}\frac{d^{2}}{dt^{2}}\left(
V_{a}-V_{b}\right) +\frac{1}{L_{0}}V_{a}  \notag \\
&+\sum_{b}\frac{1}{L_{ab}}\left( V_{a}-V_{b}\right) +C_{0}\frac{d^{2}}{dt^{2}%
}V_{a},
\end{align}%
where $I_{a}$ is the current between site $a$ and the ground, $V_{a}$ is the
voltage at site $a$, $C_{ab}$ is the capacitance and $L_{ab}$ is the
inductance between sites $a$ and $b$, and the sum is taken over all adjacent
sites $b$, while $L_{0}$ is the inductance and $C_{0}$ is the capacitance
between site $a$ and the ground.

By making the Fourier transformation, $I_{a}\left( t\right) =I_{a}\left(
\omega \right) e^{i\omega t}$ and $V_{a}\left( t\right) =V_{a}\left( \omega
\right) e^{i\omega t}$, the Kirchhoff current law leads to the formula\cite%
{ComPhys,TECNature},%
\begin{align}
I_{a}\left( \omega \right) & =\sum_{b}i\omega C_{ab}\left(
V_{a}-V_{b}\right) +\frac{1}{i\omega L_{0}}V_{a}  \notag \\
& +\sum_{b}\frac{1}{i\omega L_{ab}}\left( V_{a}-V_{b}\right) +i\omega
C_{0}V_{a},  \label{Kirch}
\end{align}%
which is summarized as%
\begin{equation}
I_{a}\left( \omega \right) =\sum_{b}J_{ab}\left( \omega \right) V_{b}\left(
\omega \right) ,  \label{CircuLap}
\end{equation}%
where the sum is taken over all adjacent sites $b$. Here, $J_{ab}\left(
\omega \right) $ is called the circuit Laplacian.

We present a detailed analysis of the circuit Laplacian $J_{ab}\left( \omega
\right) $ in Sec.\ref{SecECrevisit} as in Eq.(\ref{EqC}). We equate the
circuit Laplacian with the classical Kitaev Hamiltonian (\ref{BdG}), 
\begin{equation}
J_{ab}\left( \omega \right) =i\omega H_{ab}\left( \omega \right) .
\label{JH}
\end{equation}%
The relation between the parameters in the Kitaev model and in the electric
circuit are determined by this formula.

\begin{figure}[t]
\centerline{\includegraphics[width=0.48\textwidth]{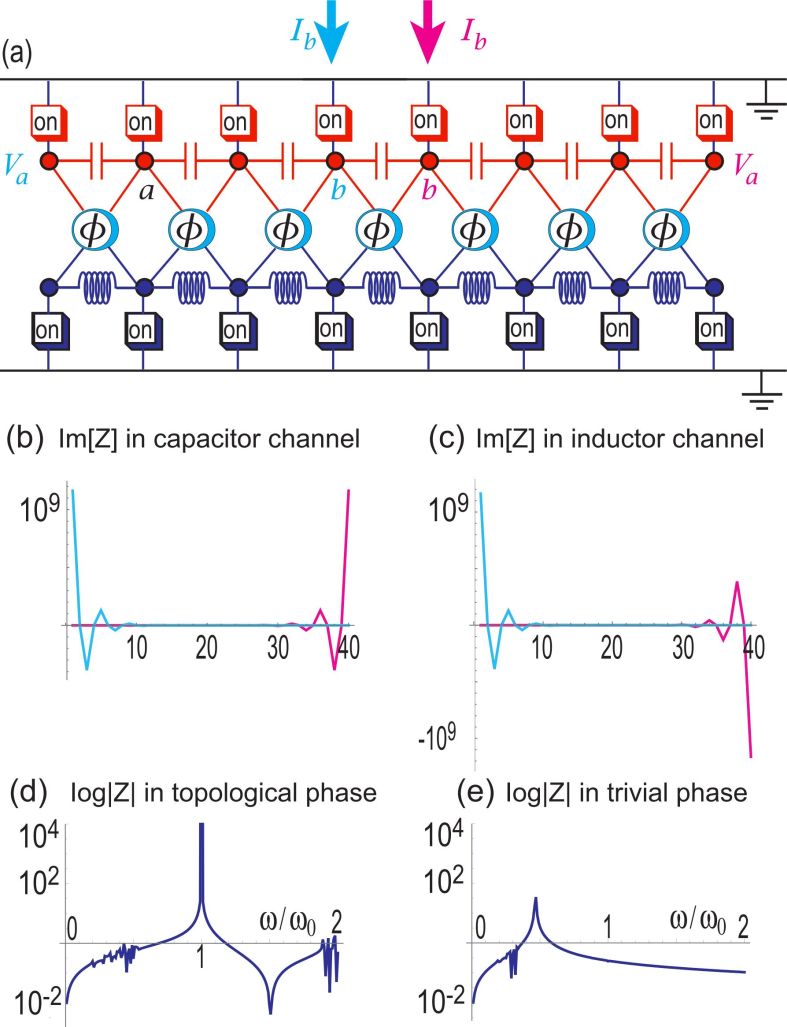}}
\caption{Kitaev chain in the topological phase. When the current is injected
at the magenta (cyan) site $b$ in (a), the impedance $Z_{ab}$ is given by a
magenta (cyan) curve as a function of the site $a$ in (b) and (c). An
impedance peak appears in Im[$Z_{ab}$] as indicated in (b) and (c), when we
take the site $a$ in the capacitor channel and the inductor channel,
respectively. The frequency is taken at the critical one $\protect\omega %
_{0} $. (d) Impedance is given as a function of $\protect\omega /\protect%
\omega _{0}$. A huge peak appears at the critical one in the topological
phase ($\protect\mu =0$). (e) There is no peak at $\protect\omega =\protect%
\omega _{0}$. in the trivial phase ($\protect\mu =4t$). We have considered a
Kitaev chain containing $40$ sites, and we have set $\Delta =0.9t$.}
\label{FigWResonance}
\end{figure}

In general, the voltage is uniquely determined in terms of the current by
the Kirchhoff law (\ref{CircuLap}). However, this is not the case for the
zero-energy sector, for which we obtain%
\begin{equation}
\sum_{b}H_{ab}\left( \omega \right) \Psi _{b}\left( \omega \right) =0,
\label{EigenEqKir}
\end{equation}%
where we have identified the voltage function,%
\begin{align}
\Psi (\omega )& =\frac{1}{\sqrt{\sum_{a}|V_{a}|^{2}}}(\cdots
,V_{0}^{e},V_{1}^{e},\cdots ,V_{m}^{e},\cdots ;  \notag \\
& \quad \quad \quad \quad \cdots ,V_{0}^{h},V_{1}^{h},\cdots
,V_{m}^{h},\cdots )  \label{WaveNodeV}
\end{align}%
as the wave function. Here, $V_{0}^{e}$ and $V_{m}^{e}$ ($V_{0}^{h}$ and $%
V_{m}^{h}$) are the voltages at the edges in the electron (hole) sector of
the topological segment $\mathcal{T}_{1}$ in the Kitaev chain in Fig.\ref%
{FigBasicBraid}.

\begin{figure}[t]
\centerline{\includegraphics[width=0.48\textwidth]{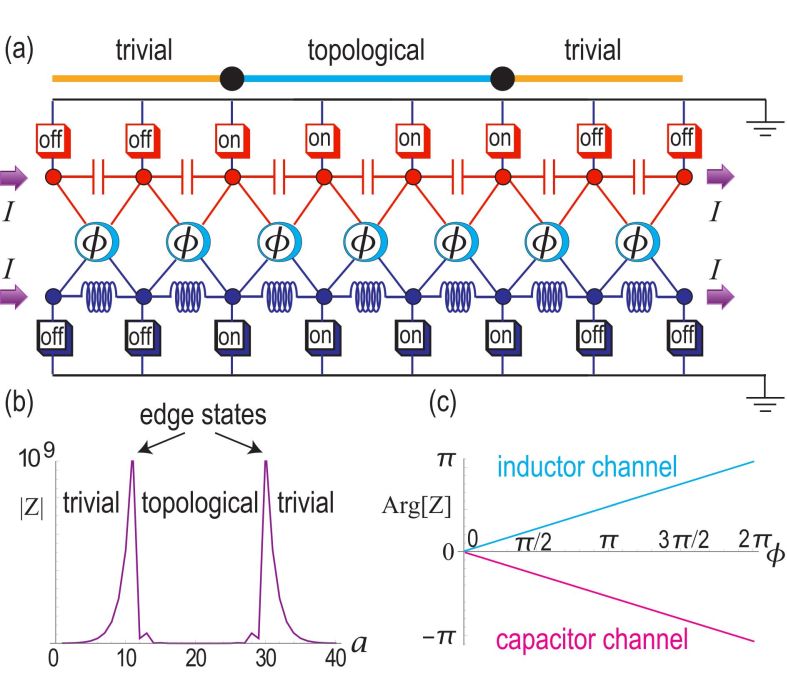}}
\caption{(a) Electric circuit for a Kitaev chain with a topological segment
and two adjacent trivial segments. (b) Absolute value of the impedance
resonance at the critical frequency $\protect\omega _{0}$, showing the
emergence of a pair of topological edge states in a Kitaev chain containing $%
40$ sites. They are located at the site $a=11$ and $a=30$. We take $\Delta
/t=0.9$ for illustration. The peaks become strictly localized for $\Delta
/t=1$. (c) $\protect\phi $ dependence of Arg$\left[ Z\right] $ for the edge
state in the capacitor channel (magenta) and the inductor channel (cyan).}
\label{FigZImpe}
\end{figure}

\subsection{Impedance peak}

The emergence of a pair of topological edge states is observed electrically
by feeding an external current to the chain. Since they are the zero-energy
eigenstate of the Kitaev Hamiltonian and since the energy corresponds to the
admittance, their emergence is observable by peaks in the impedance. The
impedance between the $a$ and $b$ sites is given by\cite{Hel} $%
Z_{ab}^{\left( 1\right) }\equiv V_{a}/I_{b}=G_{ab}$, where $G$ is the Green
function defined by the inverse of the Laplacian $J$, $G\equiv J^{-1}$.

First, we show the impedance $Z_{ab}^{\left( 1\right) }$ for a finite Kitaev
chain in a topological phase as a function of the site $a$, when a current
is injected from the site $b$ at the center of the chain, as shown in Fig.%
\ref{FigWResonance}(a). When $b$ is on the even (odd) site, the impedance
takes the maximum at the left (right) side of the chain as in Fig.\ref%
{FigWResonance}(b) and (c). We show the impedance at the edge as a function
of $\omega /\omega _{0}$ for the topological and trivial phases in Fig.\ref%
{FigWResonance}(d) and (e). There is a strong resonance at the critical
frequency at $\omega =\omega _{0}$ only in the topological phase, showing
the emergence of a zero-admittance state in the topological phase.

Next, we consider a Kitaev chain containing one topological segment
sandwiched by two trivial segments. As in Fig.\ref{FigZImpe}(a), we inject
the current $Ie^{i\omega t}$ from the left-hand side of the two main
channels and subtract it from the right-hand side. In Fig.\ref{FigZImpe}(b),
we show the impedance of a finite Kitaev chain as a function of the site $a$
in the capacitor (inductor) channel, which is calculated by%
\begin{equation}
Z_{a}=V_{a}/I=G_{a\text{L}}-G_{a\text{R}},
\end{equation}%
where L denotes the left-most site in the capacitor (inductor) channel and R
denotes the right-most site in the capacitor (inductor) channel. There are
peaks at the edges of the topological segment. The penetration depth is
longer in the trivial phases than that in the topological phase. In Fig.\ref%
{FigZImpe}(c), we show the angle of the impedance peak at the critical
frequency, which is well fitted by the lines 
\begin{equation}
\text{arg}\left[ Z_{a}\left( \phi \right) \right] =\log \left(
Z_{a}/\left\vert Z_{a}\right\vert \right) =\pm \frac{1}{2}\phi
\end{equation}%
for the capacitor channel ($-$) and the inductor channel ($+$) at an edge
site. Thus, the "superconducting" phase $\phi $ is observable in the
capacitor channel (magenta) and the inductor channel (cyan), as indicated by
the zero-energy solutions (\ref{GroundA}) of the Kitaev model.

\subsection{\textit{LC} resonator as information storage}

\label{LCR}

\begin{figure}[t]
\centerline{\includegraphics[width=0.48\textwidth]{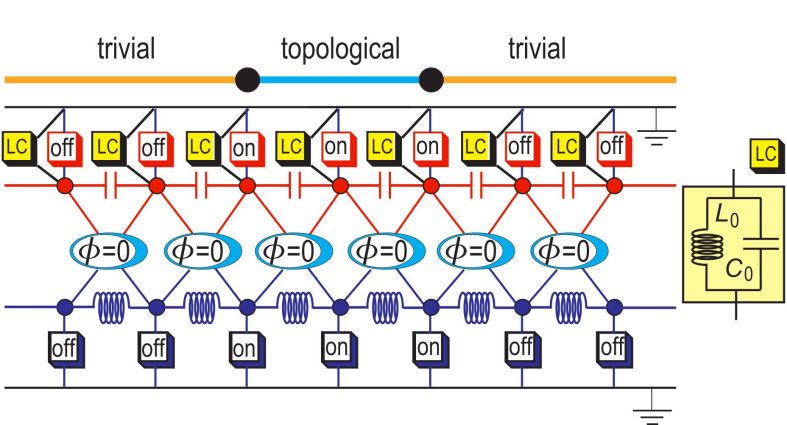}}
\caption{We attach an \textit{LC} resonator to each site to register the
phase of voltage in the capacitor channel.}
\label{FigStrage}
\end{figure}

The emergence of an edge state is observed by an impedance peak when a
current $Ie^{i\omega t}$ is injected. However, in performing a braiding
operation, we should not inject any external current since it may affect the
phase of the voltage externally. We wonder how to register and observe the
qubit information in electric circuits.

It is possible to store the qubit information in \textit{LC} resonators,
which are to be inserted to the Kitaev chain, as illustrated in Fig.\ref%
{FigStrage}. It is enough to introduce LC resonators only to the capacitor
channel, because the voltages in the two channels are related by complex
conjugation.

An \textit{LC} resonator is described by the Kirchhoff law,

\begin{equation}
I_{C}+I_{V}=0,\qquad I_{C}=C_{0}\frac{dV_{C}}{dt},\qquad V_{L}=L_{0}\frac{%
dI_{L}}{dt},
\end{equation}%
which amount to%
\begin{equation}
\frac{d^{2}I}{dt^{t}}+\omega _{0}^{2}I=0
\end{equation}%
at the critical frequency $\omega _{0}=1/\sqrt{L_{0}C_{0}}$. The solution is
given by%
\begin{equation}
I=I_{0}\sin \left( \omega _{0}t+\theta _{0}\right) ,\qquad V=V_{0}\cos
\left( \omega _{0}t+\theta _{0}\right) ,  \label{ResCos}
\end{equation}%
where $\theta _{0}$ is an initial phase at $t=0$. In the Fourier form, the
contribution of the \textit{LC} resonator to the circuit Laplacian (\ref{EqC}%
) is given by%
\begin{equation}
J=i\omega _{0}C_{0}+\frac{1}{i\omega _{0}L_{0}}=0
\end{equation}%
at the critical frequency. Hence, the edge states are not affected by the
insertion of the \textit{LC} resonator.

Furthermore, because the impedance diverges at the edge states, there is a
perfect reflection at the edge states. Thus, even when we activate an 
\textit{LC} resonator connected with an edge state, the current circularly
loops only within the \textit{LC} resonator. Nevertheless, the voltage of an
edge state is naturally the same as that of an \textit{LC} resonator.

We consider a pair of \textit{LC} resonators attached to the right and left
edges of one topological segment. Let us choose the phase of the voltage
such as $(\theta _{A},\theta _{B})=(0,0)$ or $(0,\pi )$ in the capacitor
channel. Substituting these into Eq.(\ref{ResCos}) we obtain%
\begin{align}
V(\theta _{A},\theta _{B})& =\frac{iV_{0}\cos \left( \omega _{0}t\right) }{2}
\notag \\
& \times (1,0,\cdots ,0,\left( -1\right) ^{n};-1,0,\cdots ,0,\left(
-1\right) ^{n}),  \label{V12}
\end{align}%
where $n=0$ for $(\theta _{A},\theta _{B})=(0,0)$, or $n=1$ for $(\theta
_{A},\theta _{B})=(0,\pi )$. After normalization, we obtain%
\begin{equation}
\vec{\psi}_{|n\rangle }=\frac{i}{2}(1,0,\cdots ,0,\left( -1\right)
^{n};-1,0,\cdots ,0,\left( -1\right) ^{n}),  \label{V12a}
\end{equation}%
which agrees with the wave function (\ref{EqE}) for the one-qubit states $%
|0\rangle $ and $|1\rangle $. As far as the electron sector concerns, they
are in-phase and opposite-phase states. Hence, one-qubit information can be
registered in a pair of resonators as the in-phase state $\left\vert
0\right\rangle $ or the opposite-phase state $\left\vert 1\right\rangle $.
Generalization to the system having $N$ topological segments is
straightforward.

\begin{figure*}[t]
\centerline{\includegraphics[width=0.98\textwidth]{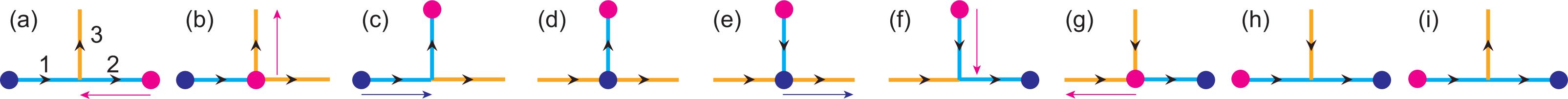}}
\caption{Illustrations of a braiding process of the two edges across a
topological segment (cyan) with the use of a T-junction. The two edges in
the initial configuration (a) are braided into those in the final
configuration (i). Arrows $\rightarrow ,\uparrow $ and $\leftarrow
,\downarrow $ on the legs represent the phase $\protect\phi =0$ and $\protect%
\phi =\protect\pi $, respectively. The phase is rotated from $\protect\phi %
=0 $ to $\protect\phi =\protect\pi $ between (d) to (e), and from $\protect%
\phi =\protect\pi $ to $\protect\phi =0$ between (h) to (i). Long arrows
beside legs indicate the direction toward which an edge is moved. }
\label{FigTBraid}
\end{figure*}

\begin{figure}[t]
\centerline{\includegraphics[width=0.48\textwidth]{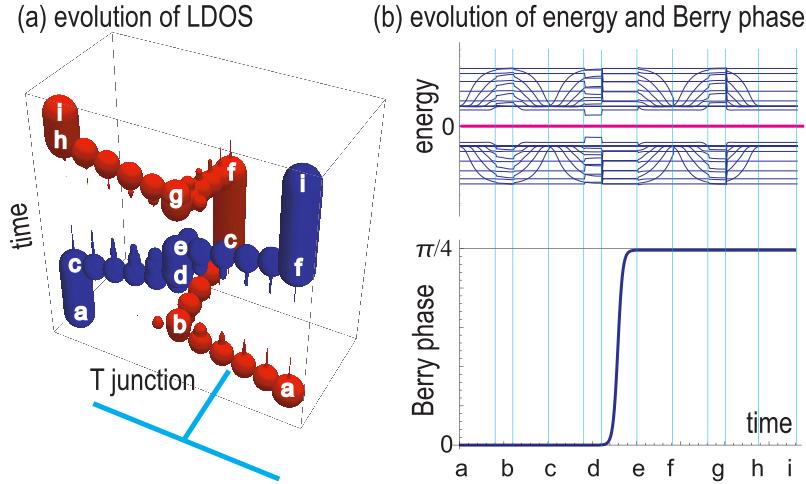}}
\caption{ (a) Evolution of the local density of states (LDOS) describing the
two edges of a topological segment. The evolution is smooth. (b) Evolution
of the energy and the Berry phase as a function of time. The zero-energy
edge states (red line) are well separated from the bulk states (blue curves)
during the braiding process. Both edges acquire the Berry phase $\protect\pi %
/4$ during the process (d) to (e) in Fig.\protect\ref{FigTBraid}. Each leg
has five sites. Alphabets a,b, $\cdots$, i denotes the steps (a), (b), $%
\cdots$, (i) of the braiding process in Fig.\protect\ref{FigTBraid}. }
\label{FigXBraid}
\end{figure}

\section{Braiding process}

\label{SecBraid}

\subsection{Berry phase}

We investigate how a set of eigenfunctions describing a pair of edge states
evolves when a system parameter is locally modified from an initial value $%
\Omega _{0}$ to another value $\Omega $.

Let $\vec{\psi}_{\beta }(\Omega )_{\text{initial}}$ be the zero-energy
eigenfunction of the Hamiltonian $H(\Omega )$,%
\begin{equation}
H(\Omega )\vec{\psi}_{\beta }(\Omega )_{\text{initial}}=0.  \label{EigenZero}
\end{equation}%
As $\Omega $ changes, it evolves as\cite%
{TQC,Kitaev,NPJ,Bonder,Cheng,Pahomi,EzawaMajo}%
\begin{equation}
\vec{\psi}_{\alpha }(\Omega )=U^{\alpha \beta }(\Omega ,\Omega _{0})\vec{\psi%
}_{\beta }(\Omega )_{\text{initial}},  \label{Wave}
\end{equation}%
where%
\begin{equation}
U(\Omega ,\Omega _{0})=e^{i\Gamma (\Omega ,\Omega _{0})},  \label{BerryU}
\end{equation}%
with $\Gamma \left( \Omega ,\Omega _{0}\right) $ the Berry phase\cite%
{Wilczek} defined by 
\begin{equation}
\Gamma _{\alpha \beta }\left( \Omega ,\Omega _{0}\right) =-i\int_{\Omega
_{0}}^{\Omega }\vec{\psi}_{\alpha }^{\dag }d\vec{\psi}_{\beta }.
\label{Berry}
\end{equation}%
Here, $\vec{\psi}_{\alpha }$ is the electron-sector part of the
eigenfunction in Eq.(\ref{GroundA}). The Berry phase accumulation is
opposite between the electron and hole sectors, as follows from the PHS.

We consider two basic examples. First, we control the phase $\phi $ locally.
Let us choose $\phi =0$ for the initial state. When it increases from $\phi
=0$ to $\phi =\Phi $, the Berry phase is%
\begin{equation}
\Gamma _{\alpha \beta }\left( \Phi \right) =-i\int_{0}^{\Phi }\vec{\psi}%
_{\alpha }^{\dag }(\phi )\partial _{\phi }\vec{\psi}_{\beta }(\phi )d\phi
=\delta _{\alpha \beta }\frac{\Phi }{4},  \label{Phase}
\end{equation}%
where we have used $\vec{\psi}_{\alpha }^{\dag }(\phi )\partial _{\phi }\vec{%
\psi}_{\beta }(\phi )=(i/4)\delta _{\alpha \beta }$.

Second, we control the length of a topological segment by tuning locally the
chemical potential $\mu $. We obtain $\Gamma _{\alpha \beta }=0$ when we
shift the position of an edge without changing the phase $\phi $. For
example, we consider the following states in the range $0\leq x\leq \pi /2$,%
\begin{align}
\vec{\psi}_{A}\left( x\right) =& \frac{1}{\sqrt{2}}\{e^{-i\phi /2}\cos
x,e^{-i\phi /2}\sin x,\cdots ,0;  \notag \\
& \qquad \qquad e^{i\phi /2}\cos x,e^{i\phi /2}\sin x,\cdots ,0\},  \notag \\
\vec{\psi}_{B}\left( x\right) =& \frac{1}{\sqrt{2}}\{0,\cdots ,-ie^{-i\phi
/2}\sin x,-ie^{-i\phi /2}\cos x;  \notag \\
& \qquad \qquad 0,\cdots ,ie^{i\phi /2}\sin x,ie^{i\phi /2}\cos x\}.
\end{align}%
They describe the edge states (\ref{GroundA}) when $x=0$. As $x$ increases
from $x=0$ to $x=\pi /2$, the edge moves just by one site. Then we find $%
\vec{\psi}_{\alpha }^{\dag }d\vec{\psi}_{\beta }=\vec{\psi}_{\alpha }^{\dag
}(x)\partial _{x}\vec{\psi}_{\beta }(x)dx=0$.

\begin{figure*}[t]
\centerline{\includegraphics[width=0.98\textwidth]{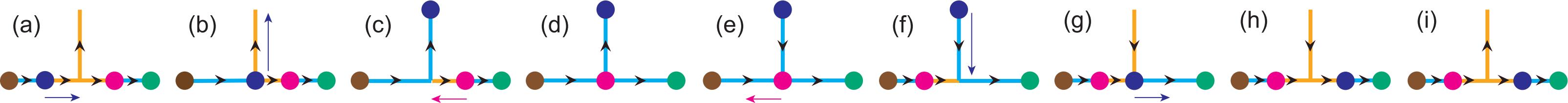}}
\caption{Illustrations of a braiding process of the two edges across a
trivial segment (orange) with the use of a T-junction. See also the caption
of Fig.\protect\ref{FigTBraid}.}
\label{Fig4Braid}
\end{figure*}

\begin{figure}[t]
\centerline{\includegraphics[width=0.48\textwidth]{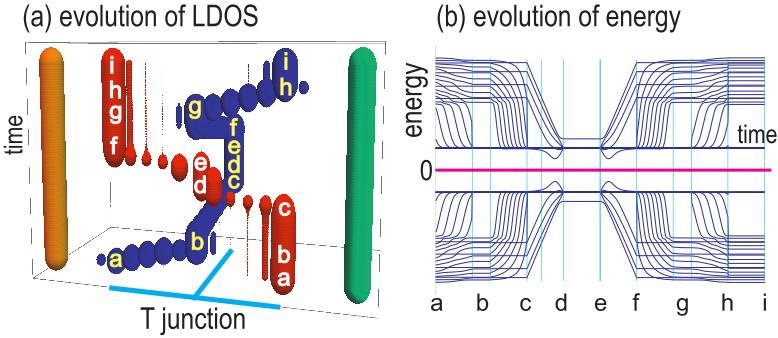}}
\caption{ (a) Evolution of the LDOS as the two edges are braided across a
trivial segment. (b) Evolution of the energy as a function of time. Each leg
has ten sites. Alphabets a,b, $\cdots $, i denotes the steps (a), (b), $%
\cdots $, (i) of the braiding process in Fig.\protect\ref{Fig4Braid}. See
also the caption of Fig.\protect\ref{FigXBraid}. }
\label{Fig4XBraid}
\end{figure}

\subsection{Braiding of two edges of a topological segment}

\label{SSecAcrossTopol}

We cannot braid two edges with the use of a single chain. This problem has
been solved by employing a T-junction\cite{AliceaBraid,AA} as shown in Fig.%
\ref{FigTBraid}. We consider a T-junction with three legs (named 1, 2, and
3) made of Kitaev chains. We set $\phi =0$ in all legs. Such a structure is
designed in electric circuits as in Fig.\ref{FigTCircuit}. We prepare the
initial state [Fig.\ref{FigTBraid}(a)], which consists of the two horizontal
legs 1 and 2 made topological and the vertical leg 3 made trivial. As we
have already stated, it is possible to make a portion of a chain topological
by controlling the chemical potential $\mu $ locally.

We braid two edges following the eight steps from (a) to (i) in the process
as shown in Fig.\ref{FigTBraid}, where the initial and final states in Fig.%
\ref{FigTBraid}(a) and (i) are identical.

We explain the eight steps from (a) to (f) in the braiding of two edges of a
topological segment in some details: See \ref{FigTBraid}.

Two edge states emerge at the left and right hands of the horizontal line as
indicated in Fig.\ref{FigTBraid}(a).

(1: a$\rightarrow $b) We move the edge on leg 2 toward the T-junction, by
making the topological segment on leg 2 shorter. This process is done as
explained in eq.(6) analytically.

(2: b$\rightarrow $c) When the edge reaches at the junction, we turn the
trivial segment on leg 2 topological gradually so that the edge moves upward.

(3: c$\rightarrow $d) When leg 3 becomes topological entirely, we move the
edge on leg 1 toward the T-junction.

(4: d$\rightarrow $e) When the edge reaches at the junction, we rotate the
phase of leg 3 from $\phi _{\text{ini}}=0$ to $\phi _{\text{fin}}=\pi $.
This process is done as explained in eq.(4) analytically.

(5: e$\rightarrow $f) When the phase of leg 3 becomes $\phi =\pi $, we move
the edge right on leg 2.

(6: f$\rightarrow $g) When leg 2 becomes topological entirely, we move the
edge on leg 3 downward.

(7: g$\rightarrow $h) When the edge on leg 3 reaches at the junction, we
move it leftward on leg 1.

(8: h$\rightarrow $i) When leg 1 becomes topological entirely, we rotate the
phase of the trivial segment on leg 3 from $\phi _{\text{ini}}=\pi $ to $%
\phi _{\text{fin}}=0$ on leg 3.

We present numerical results of this braiding process in Fig.\ref{FigXBraid}%
, which demonstrates that the process proceeds smoothly. It is confirmed
that the energy of the edge states remains zero during the process and that
the edge states are well separated from all other states as in Fig.\ref%
{FigXBraid}(b).

It is also confirmed that the Berry phase $\pi /4$ is acquired in the
process (4: d $\rightarrow $ e) in Fig.\ref{FigXBraid}(b),%
\begin{equation}
\Gamma _{\alpha \beta }(\text{d}\rightarrow \text{e})=\frac{\pi }{4}\delta
_{\alpha \beta },
\end{equation}
and it follows from (\ref{BerryU}) that%
\begin{equation}
U_{\alpha \beta }(\text{d}\rightarrow \text{e})=e^{i\pi /4}\delta _{\alpha
\beta },  \label{WA}
\end{equation}%
where the phase of a topological segment is rotated from $\phi =0$ to $\phi
=\pi $. There is no other contribution from the Berry phase.

There exists another contribution to the unitary transformation $U$ in Eq.(%
\ref{Wave}), resulting from the fact that two states $\vec{\psi}_{A}$ and $%
\vec{\psi}_{B}$ are interchanged in the process (a$\rightarrow $i). The
eigenfunctions (\ref{GroundA}) are explicitly given by 
\begin{align}
\vec{\psi}_{A}\left( 0\right) & =\frac{1}{\sqrt{2}}\left\{ i,0,\cdots
,0;-i,0,\cdots ,0\right\} ,  \notag \\
\vec{\psi}_{B}\left( 0\right) & =\frac{1}{\sqrt{2}}\left\{ 0,\cdots
,0,1;0,\cdots ,0,1\right\} ,  \label{GroundAA}
\end{align}%
at $\phi =0$, which are brought to 
\begin{align}
\vec{\psi}_{A}\left( \pi \right) & =\frac{1}{\sqrt{2}}\left\{ 1,0,\cdots
,0;1,0,\cdots ,0\right\} ,  \notag \\
\vec{\psi}_{B}\left( \pi \right) & =\frac{1}{\sqrt{2}}\left\{ 0,\cdots
,0,-i;0,\cdots ,0,i\right\} ,
\end{align}%
at $\phi =\pi $. Thus, by the phase change ($\phi :0\rightarrow \pi $) at
the step (4: d$\rightarrow $e) and by the interchange of the two edges, we
obtain%
\begin{equation}
\vec{\psi}_{A}\left( 0\right) \rightarrow \vec{\psi}_{B}\left( 0\right)
,\qquad \vec{\psi}_{B}\left( 0\right) \rightarrow -\vec{\psi}_{A}\left(
0\right) .  \tag{43}
\end{equation}%
On the other hand, the phase change ($\phi :\pi \rightarrow 0$) at the step
(8: h$\rightarrow $i) does not affect the phases of the edge states.
Combining these two effects, we obtain%
\begin{equation}
i\sigma _{y}\left( 
\begin{array}{c}
\vec{\psi}_{A}\left( 0\right)  \\ 
\vec{\psi}_{B}\left( 0\right) 
\end{array}%
\right) =\left( 
\begin{array}{c}
\vec{\psi}_{B}\left( 0\right)  \\ 
-\vec{\psi}_{A}\left( 0\right) 
\end{array}%
\right) .  \label{WC}
\end{equation}%
Thus, the effect is summarized as the operation $i\sigma _{y}$.

The transformations (\ref{WA}) and (\ref{WC}) contribute to the diagonal and
off-diagonal components of the unitary transformation (\ref{Wave}). By
combining these, the braiding is found to bring the initial state $(\vec{\psi%
}_{A},\vec{\psi}_{B})^{t}$ to the final state $U_{AB}(\vec{\psi}_{A},\vec{%
\psi}_{B})^{t}$ with%
\begin{equation}
U_{AB}=\exp \left[ \frac{i\pi }{4}\sigma _{y}\right] =\frac{1}{\sqrt{2}}%
\left( 
\begin{array}{cc}
1 & 1 \\ 
-1 & 1%
\end{array}%
\right) .  \label{U_LR}
\end{equation}%
The length of the legs can be as small as five sites since the penetration
depth of a topological edge state is zero.

It is necessary to know how the braiding acts on the initial state $%
(\left\vert 0\right\rangle ,\left\vert 1\right\rangle )$ in the qubit
representation. Recall that the two sets of the states are related by Eq.(%
\ref{UAB}). Hence, we find that the braiding operation $U_{12}$ acts on the
one-qubit state as%
\begin{align}
U_{12}\left( 
\begin{array}{c}
\vec{\psi}_{|0\rangle } \\ 
\vec{\psi}_{|1\rangle }%
\end{array}%
\right) & =U_{\text{basis}}U_{AB}U_{\text{basis}}^{-1}\left( 
\begin{array}{c}
\vec{\psi}_{A} \\ 
\vec{\psi}_{B}%
\end{array}%
\right)  \notag \\
& =\left( 
\begin{array}{cc}
e^{-i\pi /4} & 0 \\ 
0 & e^{i\pi /4}%
\end{array}%
\right) \left( 
\begin{array}{c}
\vec{\psi}_{|0\rangle } \\ 
\vec{\psi}_{|1\rangle }%
\end{array}%
\right) ,
\end{align}%
or%
\begin{equation}
U_{12}\left( 
\begin{array}{c}
\left\vert 0\right\rangle \\ 
\left\vert 1\right\rangle%
\end{array}%
\right) =\left( 
\begin{array}{cc}
e^{-i\pi /4} & 0 \\ 
0 & e^{i\pi /4}%
\end{array}%
\right) \left( 
\begin{array}{c}
\left\vert 0\right\rangle \\ 
\left\vert 1\right\rangle%
\end{array}%
\right) .  \label{U12}
\end{equation}%
The braiding process acts as a phase-shift gate in the qubit representation.
It follows that%
\begin{equation}
U_{12}^{2}=-i\sigma _{z},
\end{equation}%
and hence $U_{12}$ is proportional to the square root of the Z gate, 
\begin{equation}
U_{12}=\sqrt{-i}\sqrt{\sigma _{z}},  \label{Sz}
\end{equation}%
which is the $\sqrt{\sigma _{Z}}$ gate\cite{Les}.

\subsection{Braiding of two edges of a trivial segment}

\label{SSecAcrossTrivi}

We discuss the braiding of two edges of a trivial segment. The braiding
process is similar to the previous case: It occurs following nine steps from
(a) to (i), as illustrated in Fig.\ref{Fig4Braid}. A new configuration is
Fig.\ref{Fig4Braid}(d), where all three legs are topological. It is
understood that the point at the junction is the edge of the vertical
topological segment. A phase rotation by $\pi $ occurs on a topological
segment from Fig.\ref{Fig4Braid}(d) to (e) with a contributing to the Berry
phase, and on a trivial segment from Fig.\ref{Fig4Braid}(h) to (i) with no
contribution to the Berry phase. We show numerical results of this braiding
process in Fig.\ref{Fig4XBraid}, which demonstrates that the process
proceeds smoothly.

It is convenient to relabel the edge states as%
\begin{equation}
(\vec{\psi}_{A}^{1},\vec{\psi}_{B}^{1},\vec{\psi}_{A}^{2},\vec{\psi}%
_{B}^{2},\cdots )\rightarrow (\vec{\psi}_{1},\vec{\psi}_{2},\vec{\psi}_{3},%
\vec{\psi}_{4},\cdots ),  \label{relabel}
\end{equation}%
as shown in Fig.\ref{FigBasicBraid}. The braiding process involves two
topological segments, and hence it is a two-qubit operation. The two-qubit
basis is $\left\vert n_{1}n_{2}\right\rangle $ as given by Eq.(\ref{Qubit}),
the first qubit $n_{1}$ is formed by $\vec{\psi}_{1}$ and $\vec{\psi}_{2}$,
while the second qubit $n_{2}$ is formed by $\vec{\psi}_{3}$ and $\vec{\psi}%
_{4}$.

However, since the braiding occurs for a pair $\vec{\psi}_{2}$ and $\vec{\psi%
}_{3}$, it is convenient to consider two qubits $(\vec{\psi}_{2},\vec{\psi}%
_{3})$ and $(\vec{\psi}_{1},\vec{\psi}_{4})$. Hence, we introduce a new
two-qubit basis $|n_{+}n_{-}\}$, where $n_{+}=0,1$ is a qubit formed by $%
\vec{\psi}_{2}$ and $\vec{\psi}_{3}$, while $n_{-}=0,1$ is a qubit formed by 
$\vec{\psi}_{1}$ and $\vec{\psi}_{4}$.The transformation is given by the
fusion matrix of the Ising anyon\cite{Pachos,Stanes,Aasen},%
\begin{equation}
\left( 
\begin{array}{c}
|00\} \\ 
|11\} \\ 
|01\} \\ 
|10\}%
\end{array}%
\right) =\frac{1}{\sqrt{2}}\left( 
\begin{array}{cccc}
1 & 1 & 0 & 0 \\ 
-1 & 1 & 0 & 0 \\ 
0 & 0 & 1 & 1 \\ 
0 & 0 & -1 & 1%
\end{array}%
\right) \left( 
\begin{array}{c}
\left\vert 00\right\rangle \\ 
\left\vert 11\right\rangle \\ 
\left\vert 01\right\rangle \\ 
\left\vert 10\right\rangle%
\end{array}%
\right) .  \label{StepA}
\end{equation}%
The inverse relation reads%
\begin{equation}
\left( 
\begin{array}{c}
\left\vert 00\right\rangle \\ 
\left\vert 11\right\rangle \\ 
\left\vert 01\right\rangle \\ 
\left\vert 10\right\rangle%
\end{array}%
\right) =\frac{1}{\sqrt{2}}\left( 
\begin{array}{cccc}
1 & -1 & 0 & 0 \\ 
1 & 1 & 0 & 0 \\ 
0 & 0 & 1 & -1 \\ 
0 & 0 & 1 & 1%
\end{array}%
\right) \left( 
\begin{array}{c}
|00\} \\ 
|11\} \\ 
|01\} \\ 
|10\}%
\end{array}%
\right) .  \label{StepB}
\end{equation}%
Here, we recall that the formula (\ref{U12}) acts on $\left\vert
0\right\rangle $ and $\left\vert 1\right\rangle $ for the exchange of $\vec{%
\psi}_{1}$ and $\vec{\psi}_{2}$. Now, the corresponding exchange operator is 
$U_{23}$ for the exchange of $\vec{\psi}_{2}$ and $\vec{\psi}_{3}$. Since it
acts on $|0n_{-}\}$ and $|1n_{-}\}$ irrespective of the second component $%
n_{-}$, we obtain 
\begin{equation}
U_{23}\left( 
\begin{array}{c}
|00\} \\ 
|11\}%
\end{array}%
\right) =\left( 
\begin{array}{cc}
e^{-i\pi /4} & 0 \\ 
0 & e^{i\pi /4}%
\end{array}%
\right) \left( 
\begin{array}{c}
|00\} \\ 
|11\}%
\end{array}%
\right) ,  \label{U23}
\end{equation}%
and%
\begin{equation}
U_{23}\left( 
\begin{array}{c}
|01\} \\ 
|10\}%
\end{array}%
\right) =\left( 
\begin{array}{cc}
e^{-i\pi /4} & 0 \\ 
0 & e^{i\pi /4}%
\end{array}%
\right) \left( 
\begin{array}{c}
|01\} \\ 
|10\}%
\end{array}%
\right) .
\end{equation}%
Then, we determine how $U_{23}$ acts on the original basis. We use Eq.(\ref%
{StepB}) to find%
\begin{equation}
U_{23}\left\vert 00\right\rangle =\frac{1}{\sqrt{2}}\left(
U_{23}|00\}-U_{23}|11\}\right) .
\end{equation}%
Then, we use Eq.(\ref{U23}) to derive%
\begin{equation}
U_{23}\left\vert 00\right\rangle =\frac{1}{\sqrt{2}}\left( e^{-i\pi
/4}|00\}-e^{i\pi /4}|11\}\right) .
\end{equation}%
Finally, we use Eq.(\ref{StepA}) to find 
\begin{equation}
U_{23}\left\vert 00\right\rangle =\frac{1}{\sqrt{2}}\left( \left\vert
00\right\rangle -i\left\vert 11\right\rangle \right) .
\end{equation}%
We may carry out similar analysis for $U_{23}\left\vert 11\right\rangle $, $%
U_{23}\left\vert 01\right\rangle $ and $U_{23}\left\vert 10\right\rangle $.
The results are summarized as%
\begin{equation}
U_{23}\left( 
\begin{array}{c}
\left\vert 00\right\rangle \\ 
\left\vert 01\right\rangle \\ 
\left\vert 10\right\rangle \\ 
\left\vert 11\right\rangle%
\end{array}%
\right) =\frac{1}{\sqrt{2}}\left( 
\begin{array}{cccc}
1 & 0 & 0 & -i \\ 
0 & 1 & -i & 0 \\ 
0 & -i & 1 & 0 \\ 
-i & 0 & 0 & 1%
\end{array}%
\right) \left( 
\begin{array}{c}
\left\vert 00\right\rangle \\ 
\left\vert 01\right\rangle \\ 
\left\vert 10\right\rangle \\ 
\left\vert 11\right\rangle%
\end{array}%
\right) .  \label{EqB}
\end{equation}%
Since the parity is preserved during the braiding process, it can be
decomposed into the even parity%
\begin{equation}
U_{23}\left( 
\begin{array}{c}
\left\vert 00\right\rangle \\ 
\left\vert 11\right\rangle%
\end{array}%
\right) =\frac{1}{\sqrt{2}}\left( 
\begin{array}{cc}
1 & -i \\ 
-i & 1%
\end{array}%
\right) \left( 
\begin{array}{c}
\left\vert 00\right\rangle \\ 
\left\vert 11\right\rangle%
\end{array}%
\right) ,
\end{equation}%
and the odd parity%
\begin{equation}
U_{23}\left( 
\begin{array}{c}
\left\vert 01\right\rangle \\ 
\left\vert 10\right\rangle%
\end{array}%
\right) =\frac{1}{\sqrt{2}}\left( 
\begin{array}{cc}
1 & -i \\ 
-i & 1%
\end{array}%
\right) \left( 
\begin{array}{c}
\left\vert 01\right\rangle \\ 
\left\vert 10\right\rangle%
\end{array}%
\right) .
\end{equation}%
It follows that%
\begin{equation}
U_{23}^{2}=-i\sigma _{x}.
\end{equation}%
The operation is a square root of the NOT gate, that is the $\sqrt{\sigma
_{X}}$ gate\cite{Les},%
\begin{equation}
U_{23}=\sqrt{-i}\sqrt{\sigma _{x}}.  \label{Sx}
\end{equation}%
These results are exactly the same as those in the two-qubit operation based
on Majorana fermions\cite{Ivanov}.

\subsection{Entangled states}

We show that an entangle state is generated by a two-qubit operation. For
example, we have 
\begin{equation}
U_{23}\left\vert 00\right\rangle =\left\vert 00\right\rangle -i\left\vert
11\right\rangle .  \label{enta}
\end{equation}%
This is an entangled state. Let us prove it. If it is not, the final state
should be a pure state and written as%
\begin{align}
U_{23}\left\vert 00\right\rangle & =\left( \alpha _{1}\left\vert
0\right\rangle +\beta _{1}\left\vert 1\right\rangle \right) \otimes \left(
\alpha _{2}\left\vert 0\right\rangle +\beta _{2}\left\vert 1\right\rangle
\right)  \notag \\
& =\alpha _{1}\alpha _{2}\left\vert 00\right\rangle +\alpha _{1}\beta
_{2}\left\vert 01\right\rangle +\beta _{1}\alpha _{2}\left\vert
10\right\rangle +\beta _{1}\beta _{2}\left\vert 11\right\rangle .
\end{align}%
It follows from Eq.(\ref{enta}) that $\alpha _{1}\alpha _{2}\neq 0$ and $%
\beta _{1}\beta _{2}\neq 0$, and hence $\alpha _{1}\alpha _{2}\beta
_{1}\beta _{2}\neq 0$, which yields $\alpha _{1}\beta _{2}\neq 0$ and $\beta
_{1}\alpha _{2}\neq 0$. This contradicts Eq.(\ref{enta}). Namely, an
entangled state is produced by a braiding of two edge solitons across a
trivial segment.

\subsection{Braiding relations}

We explore the braiding relations. We have explicitly constructed $U_{12}$
and $U_{23}$. It is obvious that $U_{2j-1,2j}$ and $U_{2j,2j+1}$ have the
same expressions as these, for $j=1,2,3,\cdots $. By using the matrix
representation of these braiding operations, we can check that%
\begin{align}
& U_{j-1,j}U_{j,j+1}U_{j-1,j}=U_{j,j+1}U_{j-1,j}U_{j,j+1},  \notag \\
& U_{j,j+1}U_{j^{\prime },j^{\prime }+1}=U_{j^{\prime },j^{\prime
}+1}U_{j,j+1}\quad \text{for}\quad \left\vert j-j^{\prime }\right\vert \geq
2,  \notag \\
& U_{j-1,j}U_{j,j+1}\neq U_{j,j+1}U_{j-1,j},  \label{BraidRelC}
\end{align}%
which are the braiding relations\cite{Ivanov}.

\subsection{Clifford gates}

We have constructed the $\sqrt{\sigma _{X}}$ and $\sqrt{\sigma _{Z}}$ gates
in Eq.(\ref{Sx}) and Eq.(\ref{Sz}), respectively. The $\sqrt{\sigma _{Y}}$
gate is constructed by their successive operations as%
\begin{equation}
\sqrt{\sigma _{Y}}=\sqrt{i}\sqrt{\sigma _{X}}\sqrt{\sigma _{Z}}=\frac{1}{%
\sqrt{-i}}U_{23}U_{12}.
\end{equation}%
These sets construct the Pauli gates. On the other hand, the Hadamard gate
is constructed by\cite{Les}%
\begin{equation}
iU_{12}U_{23}U_{12}=iU_{23}U_{12}U_{23}=\frac{1}{\sqrt{2}}\left( 
\begin{array}{cccc}
1 & 0 & 0 & 1 \\ 
0 & -1 & 1 & 0 \\ 
0 & 1 & 1 & 0 \\ 
1 & 0 & 0 & -1%
\end{array}%
\right) ,
\end{equation}%
because the set of $\left\vert 00\right\rangle $ and $\left\vert
11\right\rangle $ and the set of $\left\vert 01\right\rangle $ and $%
\left\vert 10\right\rangle $ are independent in the braiding operation.

\subsection{Readout of unitary gate operation}

We have explained in Subsec.~\ref{SecIni} how we can initialize the qubit
states to the pure state $\left\vert 00\cdots 0\right\rangle $. Various
qubit states are generated by operating various unitary gate operations. The
key role is played by the Berry phase in the process of braiding, which
utilizes the gauge degree of freedom. Since the external current freezes the
gauge degree of freedom, the braiding process should be carried out in the
absence of the external current. We have explained in Subsec.~\ref{LCR} that
we can store (read) the qubit information in (from) the \textit{LC}
resonators by manipulating (measuring) the voltages at the edge states even
in such a circumstance.

In this subsection, we argue how the qubit information stored in the \textit{%
LC} resonators changes by a unitary gate operation. One qubit states are $%
\left\vert 0\right\rangle $ and $\left\vert 1\right\rangle $. The state $%
\left\vert 0\right\rangle $ ($\left\vert 1\right\rangle $) is created by
applying the same (opposite) voltage at the two edges of one topological
segment simultaneously. After the application of the voltage, the in-phase
(opposite) voltage-current oscillation starts in a set of \textit{LC}
resonators, which represent the state $\left\vert 0\right\rangle $ ($%
\left\vert 1\right\rangle $).

We discuss the one-qubit gate operation associated with the braiding of the
two edge states across a topological segment, which is represented by the
one-qubit unitary operator (\ref{U12}), or 
\begin{equation}
U_{12}|0\rangle \mapsto e^{-i\pi /4}|0\rangle ,\qquad U_{12}|1\rangle
\mapsto e^{i\pi /4}|1\rangle .
\end{equation}%
These results are observed electrically as a phase shift in the voltage at
the two edge states on the capacitor channel,%
\begin{equation}
(e^{i\omega _{0}t},e^{i\omega _{0}t})\mapsto (e^{i\omega _{0}t-i\pi
/4},e^{i\omega _{0}t-i\pi /4})
\end{equation}%
for $U_{12}|0\rangle $, and%
\begin{equation}
(e^{i\omega _{0}t},-e^{i\omega _{0}t})\mapsto (e^{i\omega _{0}t+i\pi
/4},-e^{i\omega _{0}t+i\pi /4})
\end{equation}%
for $U_{12}|1\rangle $.

We may similarly discuss the two-qubit operation associated with the
braiding of the two edge states across a trivial segment. The actual
braiding is represented by the one-qubit unitary operator (\ref{EqB}), or 
\begin{equation}
U_{23}|00\rangle \mapsto \frac{1}{\sqrt{2}}\left( 00\rangle -i\left\vert
11\right\rangle \right) .
\end{equation}%
The result of the operation $U_{23}$ on the state $|00\rangle $ is observed
electrically as a phase shift in the voltage at the four edge states on the
capacitor channel,%
\begin{align}
& (e^{i\omega _{0}t},e^{i\omega _{0}t};e^{i\omega _{0}t},e^{i\omega _{0}t}) 
\notag \\
& \quad \mapsto \frac{1}{\sqrt{2}}(e^{i\omega _{0}t},e^{i\omega
_{0}t};e^{i\omega _{0}t},e^{i\omega _{0}t})  \notag \\
& \quad \quad \quad \quad -\frac{i}{\sqrt{2}}(e^{i\omega _{0}t},-e^{i\omega
_{0}t};e^{i\omega _{0}t},-e^{i\omega _{0}t})  \notag \\
& =(e^{i\omega _{0}t-i\pi /4},e^{i\omega _{0}t+i\pi /4};e^{i\omega
_{0}t-i\pi /4},e^{i\omega _{0}t+i\pi /4}).
\end{align}%
Similar results are obtained for the other states. In general, the change of
phase shift is registered in a set of \textit{LC} resonators in any gate
processing. We can read the qubit information by measuring the phase shifts
from a set of resonators after all braidings are over.

\section{Creation and annihilation of edge solitons}

\label{SecMajorana}

We have derived the basic formulas familiar in topological quantum
computations by calculating the Berry phase with the use of T-junction
geometry in the electric-circuit formalism. It is interesting to reformulate
our scheme to know to what extent we can simulate the standard
Majorana-operator formalism by the electric-circuit formalism.

A topological edge state emerges at the boundary between the topological and
trivial segments of a Kitaev chain and is observable as an impedance peak in
electric circuit. Such a state is interpreted as a topological soliton
because it is a localized particle-like object and has a topological
stability by intertwining the topological and the trivial segments. We call
it an edge soliton when we focus on the aspect of soliton. An edge soliton
in the present model is quite similar to a sine-Gordon soliton in the
sine-Gordon model. Exchange statistic of topological solitons is intriguing:
A topological soliton in the classical sine-Gordon model has been argued\cite%
{Coleman,Mandel} to be a Thirring fermion\cite{Thirring}. Similarly, we now
argue that an edge soliton in the Kitaev model behaves as a Majorana-fermion
because the circuit contains the capacitor channel and the inductor channel
corresponding to the electron band and the hole band.

We have obtained two zero-energy solutions (\ref{GroundA}), where $\vec{\psi}%
_{A}$ and $\vec{\psi}_{B}$ are perfectly localized at the left edge and the
right edge of a topological segment, respectively. They describe a pair of
edge solitons. As we have remarked in Eq.(\ref{JH}), the Kitaev Hamiltonian
is equivalent to the circuit Laplacian. The two systems are equivalent at
the Hamiltonian level, and there is one-to-one correspondence between the
wave functions in these two systems. The wave function is the voltage
function (\ref{WaveNodeV}) in electric circuits. Edge solitons are
materialized as impedance peaks in the electric circuit: See Figs.\ref%
{FigWResonance} and \ref{FigZImpe}.

Here, let us summarize key features of edge solitons.

(i) We start with the electric circuit with all TCUs being off, which
describes the Kitaev chain in the trivial phase.

(ii) We create $N$ topological segments together with $N$ pairs of edge
solitons by switching on the TCUs attached to those segments.

(ii) A pair of edge solitons are observable by peaks in voltage or impedance
both in the capacitor channel and the inductor channel: See Fig.\ref%
{FigZImpe}(c).

(iii) Two edge solitons cannot occupy a single site, which means that they
are subject to the exclusion principle.

(iv) We may move an edge freely by expanding or shrinking a topological
segment by switching on or off TCUs, which means that we can flit an edge
soliton from one site to a neighboring site.

These properties of edge solitons allow us to introduce a creation operator
of one edge soliton. The wave function $\vec{\psi}_{A}$ in Eqs.(\ref{GroundA}%
) implies that one edge soliton has a component with phase $ie^{-i\phi /2}$
in the electron sector and a portion of phase $-ie^{i\phi /2}$ in the hole
sector at $x=0$. It describes a creation of an impedance peak carrying the
corresponding phases in the capacitor channel and the inductor channel at
site $x=0 $. A similar creation operator is introduced with respect to the
edge soliton with the wave function $\vec{\psi}_{B}$ at $x=m$.

\subsection{One topological segment}

Since an appropriately designed electric circuit is equivalent to the Kitaev
model at the Hamiltonian level, we may reformulate such an electric circuit
as a lattice model. We recall that an edge soliton is observable as an
impedance peak, which consists of two parts in the capacitor channel and
inductor channel as in Fig.\ref{FigZImpe}(c). Let us define the creation
operator of an edge soliton with phase $\phi =0$ in the capacitor (inductor)
channel at $x$ by $a_{x}^{\dagger }$ ($b_{x}^{\dagger }$). It must be that $%
a_{x}^{\dagger 2}=b_{x}^{\dagger 2}=0$ due to the exclusion principle
imposed on the edge soliton in each channel.

We then define the creation operators $\gamma _{A}^{\dagger }$ and $\gamma
_{B}^{\dagger }$ of the edge solitons with phase $\phi $ by%
\begin{align}
\gamma _{A}^{\dagger }& =\sqrt{2}\vec{a}^{\dagger }\cdot \vec{\psi}%
_{A}=ie^{-i\phi /2}a_{0}^{\dagger }-ie^{i\phi /2}b_{0}^{\dagger },  \notag \\
\gamma _{B}^{\dagger }& =\sqrt{2}\vec{a}^{\dagger }\cdot \vec{\psi}%
_{B}=e^{-i\phi /2}a_{m}^{\dagger }+e^{i\phi /2}b_{m}^{\dagger },
\label{GammaPre}
\end{align}%
where $\vec{\psi}_{A}$ and $\vec{\psi}_{B}$ are the wave functions given by
Eqs.(\ref{GroundA}), and 
\begin{equation}
\vec{a}^{\dagger }=(a_{0}^{\dagger },a_{1}^{\dagger },\cdots ,a_{m}^{\dagger
};b_{0}^{\dagger },b_{1}^{\dagger },\cdots ,b_{m}^{\dagger }).  \label{OpeAB}
\end{equation}%
Although it appears that $\gamma _{A}$ is composed of $2(m+1)$ independent
components, this is not the case. The electric circuit is designed to
simulate the Kitaev model (\ref{BdG}) precisely at the Hamiltonian level,
which originally contains the same information in the electron sector and
the hole sector related via complex conjugation. Indeed, the wave functions $%
\vec{\psi}_{A}$ and $\vec{\psi}_{B}$ described by Eqs.(\ref{GroundA}) have
this property. Consequently it should be that $b_{j}^{\dagger }=a_{j}$, or%
\begin{equation}
\vec{a}^{\dagger }=(a_{0}^{\dagger },a_{1}^{\dagger },\cdots ,a_{m}^{\dagger
};a_{0},a_{1},\cdots ,a_{m}).  \label{OpeA}
\end{equation}%
Then, it follows from Eqs.(\ref{GammaPre}) that $\gamma _{A}=\gamma
_{A}^{\dagger }$ and $\gamma _{B}=\gamma _{B}^{\dagger }$. We rewrite Eqs.(%
\ref{GammaPre}) as%
\begin{align}
\gamma _{A}& =ie^{-i\phi /2}a_{0}^{\dagger }-ie^{i\phi /2}a_{0},  \notag \\
\gamma _{B}& =e^{-i\phi /2}a_{m}^{\dagger }+e^{i\phi /2}a_{m}.
\label{GammaA}
\end{align}

Now, the exclusion principle indicates that there are only two states at one
site, i.e., whether an edge soliton is absent or present at one site, which
we denote $|0\rangle \!\rangle $ and $|1\rangle \!\rangle $, with $%
a_{x}^{\dagger }|0\rangle \!\rangle =|1\rangle \!\rangle $ and $%
a_{x}|1\rangle \!\rangle =|0\rangle \!\rangle $. In the matrix form, these
relations are written in the form of%
\begin{align}
a_{x}^{\dagger }\left( 
\begin{array}{c}
|0\rangle \!\rangle \\ 
|1\rangle \!\rangle%
\end{array}%
\right) &=\left( 
\begin{array}{cc}
0 & 1 \\ 
0 & 0%
\end{array}%
\right) \left( 
\begin{array}{c}
|0\rangle \!\rangle \\ 
|1\rangle \!\rangle%
\end{array}%
\right) , \\
a_{x}\left( 
\begin{array}{c}
|0\rangle \!\rangle \\ 
|1\rangle \!\rangle%
\end{array}%
\right) &=\left( 
\begin{array}{cc}
0 & 0 \\ 
1 & 0%
\end{array}%
\right) \left( 
\begin{array}{c}
|0\rangle \!\rangle \\ 
|1\rangle \!\rangle%
\end{array}%
\right) ,
\end{align}%
which lead to%
\begin{equation}
a_{x}^{\dagger }=\left( 
\begin{array}{cc}
0 & 1 \\ 
0 & 0%
\end{array}%
\right) ,\qquad a_{x}=\left( 
\begin{array}{cc}
0 & 0 \\ 
1 & 0%
\end{array}%
\right) .
\end{equation}%
We obtain the anticommutation relation, $a_{x}a_{x}^{\dagger
}+a_{x}^{\dagger }a_{x}=1$, for an edge soliton in each channel, showing
that it behaves as a fermion. Consequently, we obtain $\gamma
_{A}^{2}=\gamma _{B}^{2}=1$.

However, we have no information on the commutation relation between $\gamma
_{A}$ and $\gamma _{B}$, which is the exchange statistics of edge solitons.

\subsection{Many topological segments}

We next study a Kitaev chain containing $N$ topological segments. For
definiteness each topological (trivial) segment is assumed to be made of $%
m+1 $ ($m-1$) sites, as in Fig.\ref{FigBasicBraid}. The $j$-th topological
segment produces two topological edge states $\vec{\psi}_{A}^{j}$ and $\vec{%
\psi}_{B}^{j}$ given by Eqs.(\ref{GroundA}). Then, we may introduce a set of
the operators as 
\begin{equation}
\gamma _{A}^{j}=\sqrt{2}\vec{a}_{j}^{\dagger }\cdot \vec{\psi}_{A},\qquad
\gamma _{B}^{j}=\sqrt{2}\vec{a}_{j}^{\dagger }\cdot \vec{\psi}_{B}
\label{ManyGamma}
\end{equation}%
with%
\begin{equation}
\vec{a}_{j}^{\dagger }=(a_{j^{\prime }}^{\dagger },a_{j^{\prime
}+1}^{\dagger },\cdots ,a_{j^{\prime }+m}^{\dagger };a_{j^{\prime
}},a_{j^{\prime }+1},\cdots ,a_{j^{\prime }+m}),  \label{ManyOpeA}
\end{equation}%
where $j^{\prime }=(2j-2)m$.

It may be convenient to relabel the edge states as%
\begin{equation}
(\gamma _{A}^{1},\gamma _{B}^{1},\gamma _{A}^{2},\gamma _{B}^{2},\cdots
)\rightarrow (\gamma _{1},\gamma _{2},\gamma _{3},\gamma _{4},\cdots )
\label{relabelG}
\end{equation}%
in accord with Eq.(\ref{relabel}). Then, it follows from the above arguments
that 
\begin{equation}
(\gamma _{j})^{2}=1,  \label{Gamma21}
\end{equation}%
for any $j$, because any edge soliton has components both in the electron
sector and the hole sector, and each component is subject to the exclusion
principle. We shall see that $\gamma _{j}$ is the Majorana operator in the
succeeding section.

\section{Exchange statistics}

We investigate the exchange statistics of edge solitons. We have already
studied the way and the result of exchanging two edge states by calculating
the Berry phase with the use of T-junction geometry in Sec.~\ref{SecBraid}.
Here we reanalyze the problem in terms of edge soliton operators. We
consider a T-junction with three legs (named 1, 2, and 3) made of Kitaev
chains as shown in Fig.\ref{FigTBraid} and Fig.\ref{Fig4Braid}. We set $\phi
=0$ in all legs.

\subsection{Braiding of two edges across a topological segment}

We braid two edge solitons across a topological segment, following the eight
steps from (a) to (i) in Fig.\ref{FigTBraid}, which we have studied in
Subsec.~\ref{SSecAcrossTopol}. We examine how these processes affect the
edge solitons at $x=1$ and $x=m$. There are two effects. (i) If the edge
soliton $\gamma _{A}$ ($\gamma _{B}$) at $x=0$ ($m$) were brought to $x=m$ ($%
0$) without changing the "superconducting phase" $\phi $, we would have

\begin{align}
\gamma _{A}& =ia_{0}^{\dagger }-ia_{0}\rightarrow ia_{m}^{\dagger }-ia_{m}, 
\notag \\
\gamma _{B}& =a_{m}^{\dagger }+a_{m}\rightarrow a_{0}^{\dagger }+a_{0},
\end{align}%
by setting $\phi =0$ in Eqs.(\ref{GammaA}), and then by exchanging the
indices $0$ and $m$. (ii) Actually, by the phase change at the step (4: d$%
\rightarrow $e), the eigenfunctions (\ref{GroundA}) read%
\begin{equation}
\gamma _{A}=a_{0}^{\dagger }+a_{0},\qquad \gamma _{B}=-ia_{m}^{\dagger
}+ia_{m},
\end{equation}%
by setting $\phi =\pi $ in Eqs.(\ref{GammaA}). Note that the phase change at
the step (8: h$\rightarrow $i) does not affect the edge solitons. Combining
these two effects, we obtain 
\begin{equation}
\gamma _{A}\rightarrow -\gamma _{B},\qquad \gamma _{B}=\gamma _{A},
\label{WBa}
\end{equation}%
or%
\begin{equation}
\gamma _{A}\gamma _{B}\rightarrow -\gamma _{B}\gamma _{A}.  \label{WB}
\end{equation}%
This is the result of a single exchange of two edge solitons in a
topological segment. A double exchange implies $\gamma _{A}\rightarrow
-\gamma _{A}$, $\gamma _{B}=-\gamma _{B}$, as expected. An edge soliton is
an Ising anyon.

\subsection{Braiding of two edges across a trivial segment}

We discuss the braiding of two edge solitons across a trivial segment,
following the eight steps from (a) to (i) in Fig.\ref{Fig4Braid}, which we
have studied in Subsec.~\ref{SSecAcrossTrivi}. We examine how these
processes affect the edge solitons at $x=m$ and $x=2m$. A phase rotation by $%
\pi $ occurs on a topological segment from Fig.\ref{Fig4Braid}(d) to (e)
with a contributing to the wave function, and on a trivial segment from Fig.%
\ref{Fig4Braid}(h) to (i) with no contribution.

We examine how these processes affect the edge solitons. For definiteness we
study the exchange of $\gamma _{B}^{1}$ and $\gamma _{A}^{2}$ in Fig.\ref%
{FigBasicBraid}, or 
\begin{align}
\gamma _{B}^{1}& =e^{-i\phi /2}a_{m}^{\dagger }+e^{i\phi /2}a_{m}=\sqrt{2}%
\vec{a}_{1}^{\dagger }\cdot \vec{\psi}_{B},  \notag \\
\gamma _{A}^{2}& =ie^{-i\phi /2}a_{2m}^{\dagger }-ie^{i\phi /2}a_{2m}=\sqrt{2%
}\vec{a}_{2}^{\dagger }\cdot \vec{\psi}_{A},  \label{GammaA2}
\end{align}%
with%
\begin{align}
\vec{a}_{1}^{\dagger }& =(a_{0}^{\dagger },a_{1}^{\dagger },\cdots
,a_{m}^{\dagger };a_{0},a_{1},\cdots ,a_{m}),  \notag \\
\vec{a}_{2}^{\dagger }& =(a_{2m}^{\dagger },a_{2m+1}^{\dagger },\cdots
,a_{3m}^{\dagger };a_{2m},a_{2m+1},\cdots ,a_{3m}).
\end{align}%
There are two effects. (i) The edge soliton $\gamma _{B}^{1}$ ($\gamma
_{A}^{2}$) at $x=m$ ($2m$) is brought to $x=2m$ ($m$), which results in%
\begin{align}
\gamma _{B}^{1}& =a_{m}^{\dagger }+a_{m}\rightarrow a_{2m}^{\dagger }+a_{2m},
\notag \\
\gamma _{A}^{2}& =ia_{2m}^{\dagger }-ia_{2m}\rightarrow ia_{m}^{\dagger
}-ia_{m},
\end{align}%
by setting $\phi =0$ in Eqs.(\ref{GammaA2}), and then by exchanging indices $%
m$ and $2m$. (ii) By the phase change at the step (4: d$\rightarrow $e), the
eigenfunctions (\ref{GroundA}) read%
\begin{equation}
\gamma _{B}^{1}=-ia_{m}^{\dagger }+ia_{m},\qquad \gamma
_{A}^{2}=a_{2m}^{\dagger }+a_{2m},
\end{equation}%
by setting $\phi =\pi $ in Eqs.(\ref{GammaA}). Note that the phase change at
the step (8: h$\rightarrow $i) does not affect the edge solitons. Combining
these two effects, we obtain 
\begin{equation}
\gamma _{B}^{1}\rightarrow -\gamma _{A}^{2},\qquad \gamma
_{A}^{2}\rightarrow \gamma _{B}^{1},  \label{WBb}
\end{equation}%
or%
\begin{equation}
\gamma _{B}^{1}\gamma _{A}^{2}\rightarrow -\gamma _{A}^{2}\gamma _{B}^{1}.
\label{WBx}
\end{equation}%
This is the result of a single exchange of two edge solitons across a
topological segment.

\subsection{Braiding operators}

When we label the edge solitons as in Eq.(\ref{relabelG}), we obtain from
Eq.(\ref{WB}) and Eq.(\ref{WBx}) that $\gamma _{j}\gamma _{j+1}=-\gamma
^{j+1}\gamma _{j}$. By repeating the exchange of adjacent edge solitons, it
is possible to generalize this result to $\gamma _{i}\gamma _{j}=-\gamma
_{j}\gamma _{i}$, which we combine with Eq.(\ref{Gamma21}) to obtain 
\begin{equation}
\gamma _{i}\gamma _{j}+\gamma _{j}\gamma _{i}=2\delta _{ij}
\end{equation}%
for any $i$ and $j$. Furthermore, the braiding results (\ref{WBa}) and (\ref%
{WBb}) are generalized to the braiding operations%
\begin{equation}
U_{ij}:\gamma _{i}\rightarrow -\gamma _{j},\qquad U_{ij}:\gamma
_{j}\rightarrow \gamma _{i},
\end{equation}%
or%
\begin{equation}
U_{ij}\gamma _{i}U_{ij}^{-1}=\gamma _{j},\qquad U_{ij}\gamma
_{j}U_{ij}^{-1}=-\gamma _{i}.
\end{equation}%
Hence, the braiding operator is given by%
\begin{equation}
U_{ij}=\frac{1}{\sqrt{2}}(1+\gamma _{j}\gamma _{i})=\exp [\pi \gamma
_{j}\gamma _{i}/4],  \label{Uij}
\end{equation}%
as agrees with the standard result\cite{Ivanov}.

\subsection{Qubits in operator formalism}

\subsubsection{One-qubit state}

We have associated the operators $\gamma _{A}$ and $\gamma _{B}$ to the edge
solitons so that they create the wave functions $\vec{\psi}_{A}$ and $\vec{%
\psi}_{B}$ given by Eqs.(\ref{GroundA}). Similarly, we may associate the
operators $f$ and $f^{^{\dagger }}$ to the states described by the wave
functions $\vec{\psi}_{|0\rangle }$ and $\vec{\psi}_{|1\rangle }$ as 
\begin{align}
f& =\vec{a}^{\dagger }\cdot \vec{\psi}_{|0\rangle }=(\gamma _{A}+i\gamma
_{B})/2,  \notag \\
f^{\dagger }& =\vec{a}^{\dagger }\cdot \vec{\psi}_{|1\rangle }=(\gamma
_{A}-i\gamma _{B})/2,  \label{f-ope}
\end{align}%
where $\vec{a}^{\dagger }$ is defined by Eq.(\ref{OpeA}). We then have%
\begin{equation}
\gamma _{A}=f+f^{\dagger },\qquad \gamma _{B}=i(f^{\dagger }-f).
\end{equation}%
The one-qubit states, $|0\rangle $ and $|1\rangle $, are defined as the
empty and the occupied states with respect to fermion operator $f$,%
\begin{equation}
f|0\rangle =0,\quad f^{\dagger }|1\rangle =0,\quad f^{\dagger }|0\rangle
=|1\rangle ,\quad f|1\rangle =|0\rangle .  \label{f-state}
\end{equation}%
The set of states $\left\vert 0\right\rangle $ and $\left\vert
1\right\rangle $ constitutes one qubit for application to topological
quantum computers.

We recall that the braiding of two edge solitons across a topological
segment is given by $U_{12}=(1+\gamma ^{2}\gamma ^{1})/\sqrt{2}$ as in Eq.(%
\ref{Uij}), or%
\begin{equation}
U_{12}=\frac{1}{\sqrt{2}}[1+i(f_{1}^{\dagger }-f_{1})(f_{1}+f_{1}^{\dagger
})].
\end{equation}%
By operating this to one-qubit states, we find%
\begin{equation}
U_{12}\left\vert 0\right\rangle =\frac{1}{\sqrt{2}}(1-i)\left\vert
0\right\rangle ,\qquad U_{12}\left\vert 1\right\rangle =\frac{1}{\sqrt{2}}%
(1+i)\left\vert 1\right\rangle .
\end{equation}%
Hence, we obtain 
\begin{equation}
U_{12}\left( 
\begin{array}{c}
\left\vert 0\right\rangle \\ 
\left\vert 1\right\rangle%
\end{array}%
\right) =\left( 
\begin{array}{cc}
e^{-i\pi /4} & 0 \\ 
0 & e^{i\pi /4}%
\end{array}%
\right) \left( 
\begin{array}{c}
\left\vert 0\right\rangle \\ 
\left\vert 1\right\rangle%
\end{array}%
\right) ,
\end{equation}%
which is nothing but Eq.(\ref{U12}).

\subsubsection{Multi-qubit state}

One-qubit states $\left\vert 0\right\rangle _{j}$ and $\left\vert
1\right\rangle _{j}$ are similarly constructed for the $j$-th topological
segment with the use of 
\begin{equation}
f_{j}=(\gamma _{A}^{j}+i\gamma _{B}^{j})/2,\quad f_{j}^{\dagger }=(\gamma
_{A}^{j}-i\gamma _{B}^{j})/2,
\end{equation}%
as in Eq.(\ref{f-state}), or%
\begin{equation}
f_{j}|1\rangle _{j}=|0\rangle _{j},\qquad f_{j}^{\dagger }|0\rangle
_{j}=|1\rangle _{j}.
\end{equation}%
Two-qubit states are defined by%
\begin{align}
& f_{1}|00\rangle =f_{2}|00\rangle =f_{1}^{\dag }|11\rangle =f_{2}^{\dag
}|11\rangle =0,  \notag \\
& |10\rangle =f_{1}^{\dag }|00\rangle ,\quad |01\rangle =f_{2}^{\dag
}|00\rangle ,\quad |11\rangle =f_{1}^{\dag }f_{2}^{\dag }|00\rangle .
\end{align}%
The many-body states are given by the direct product,%
\begin{equation}
\left\vert n_{1}n_{2}\cdots n_{N}\right\rangle =\left\vert
n_{1}\right\rangle _{1}\otimes \left\vert n_{2}\right\rangle _{2}\otimes
\cdots \otimes \left\vert n_{N}\right\rangle _{N}  \label{ManyBody}
\end{equation}%
with $n_{j}=0,1$.

We study the two-qubit states some in detail. The braiding across a trivial
segment is given by $U_{23}=(1+\gamma ^{3}\gamma ^{2})/\sqrt{2}$ as in Eq.(%
\ref{Uij}), or%
\begin{equation}
U_{23}=\frac{1}{\sqrt{2}}[1+i(f_{2}+f_{2}^{\dagger })(f_{1}^{\dagger
}-f_{1})].
\end{equation}%
By operating this to one-qubit states, we find%
\begin{equation}
U_{23}\left( 
\begin{array}{c}
\left\vert 00\right\rangle \\ 
\left\vert 01\right\rangle \\ 
\left\vert 10\right\rangle \\ 
\left\vert 11\right\rangle%
\end{array}%
\right) =\frac{1}{\sqrt{2}}\left( 
\begin{array}{cccc}
1 & 0 & 0 & -i \\ 
0 & 1 & -i & 0 \\ 
0 & -i & 1 & 0 \\ 
-i & 0 & 0 & 1%
\end{array}%
\right) \left( 
\begin{array}{c}
\left\vert 00\right\rangle \\ 
\left\vert 01\right\rangle \\ 
\left\vert 10\right\rangle \\ 
\left\vert 11\right\rangle%
\end{array}%
\right) ,
\end{equation}%
which is nothing but Eq.(\ref{EqB}).

\section{Electric-circuit realization (revisited)}

\label{SecECrevisit}

We have explained how to simulate the Kitaev chain by an electric circuit in
Sec.~\ref{SecEC}. In this section we construct the circuit Laplacians
explicitly for the topological and trivial phases.

\subsection{Topological and trivial phases}

The Kirchhoff current law (\ref{Kirch}) is summarized as Eq.(\ref{CircuLap}%
), or%
\begin{equation}
I_{a}\left( \omega \right) =\sum_{b}J_{ab}\left( \omega \right) V_{b}\left(
\omega \right) ,
\end{equation}%
where the circuit Laplacian $J_{ab}\left( \omega \right) $ is expressed as%
\begin{equation}
J=\left( 
\begin{array}{cc}
h_{1} & g_{1} \\ 
g_{2} & h_{2}%
\end{array}%
\right) .  \label{EqC}
\end{equation}%
We study $J_{ab}$ explicitly in what follows.

We first study the circuit given in the right-hand side of Fig.\ref%
{FigCircuitTT}(b). We shall show that it describes the topological phase.
Analyzing the Kirchhoff current law for the circuit, we obtain\cite%
{EzawaMajo}%
\begin{align}
h_{1}& =-2C\cos k+2C-(\omega ^{2}L_{0})^{-1},  \notag \\
h_{2}& =2(\omega ^{2}L)^{-1}\cos k-2(\omega ^{2}L)^{-1}+C_{0},
\label{TopoL-f}
\end{align}%
and%
\begin{align}
g_{1}& =-C_{X}e^{ik}+(\omega ^{2}L_{X})^{-1}e^{-ik},  \notag \\
g_{2}& =(\omega ^{2}L_{X})^{-1}e^{ik}-C_{X}e^{-ik}.  \label{TopoL-g}
\end{align}%
We make the following observation. (i) Capacitors and inductors on the main
two channels appear in the diagonal elements $h_{1}$ and $h_{2}$. (ii) Those
attached to the ground appear also appear in $h_{1}$ and $h_{2}$. (iii)
Those in the pairing interactions appear in the off-diagonal elements $g_{1}$
and $g_{2}$.

In identifying the circuit Laplacian with the Kitaev Hamiltonian as in Eq.(%
\ref{JH}), it is necessary to require 
\begin{equation}
\omega _{0}\equiv 1/\sqrt{LC}=1/\sqrt{L_{0}C_{0}}=1/\sqrt{L_{X}C_{X}}
\label{OmegaLC}
\end{equation}%
for PHS to hold for the circuit. At $\omega =\omega _{0}$, the circuit
Laplacian (\ref{EqC}) is reduced to 
\begin{equation}
J=\left[ 2C(1-\cos k)-C_{0}\right] \sigma _{z}+2C_{X}\sigma _{y}\sin k.
\label{MatrixJ}
\end{equation}%
It follows from Eq.(\ref{JH}) that 
\begin{equation}
t=-C,\quad \mu =-2C+C_{0},\quad \Delta =C_{X}.  \label{EqA}
\end{equation}%
The parameters charactering the Kitaev model (\ref{BdG}) are determined by
these equations in terms of electric elements for the right-hand side of Fig.%
\ref{FigCircuitTT}(b). The system is topological since $\left\vert \mu
\right\vert <\left\vert 2t\right\vert $ is satisfied.

Next, we study the critical point. When the capacitors $C_{0}$ and the
inductors $L_{0}$ connected to the ground are removed, the circuit Laplacian
reads%
\begin{equation}
J=2C(1-\cos k)\sigma _{z}+2C_{X}\sigma _{y}\sin k,
\end{equation}%
which is given by setting $C_{0}=0$ in Eq.(\ref{MatrixJ}). Then, Eqs.(\ref%
{TopoL-f}) are modified as 
\begin{align}
h_{1}& =-2C\cos k+2C,  \notag \\
h_{2}& =2(\omega _{0}^{2}L)^{-1}\cos k-2(\omega _{0}^{2}L)^{-1},
\end{align}%
by setting $C_{0}=0$ and $L_{0}\rightarrow \infty $. Then, the chemical
potential is given by 
\begin{equation}
\mu =-2C.
\end{equation}%
The system is precisely at the topological phase-transition point $%
\left\vert \mu \right\vert =\left\vert 2t\right\vert $, since the condition $%
\mu =2t$ is satisfied.

Finally, we study the circuit given in the left-hand side of Fig.\ref%
{FigCircuitTT}(b), which is obtained by interchanging $C_{0}$ and $L_{0}$ in
the right-hand side of the same figure. The circuit Laplacian is given by%
\begin{equation}
J=\left[ 2C(1-\cos k)+C_{0}\right] \sigma _{z}+2C_{X}\sigma _{y}\sin k,
\end{equation}%
instead of Eqs.(\ref{MatrixJ}), and we obtain%
\begin{align}
h_{1}& =-2C\cos k+2C+C_{0},  \notag \\
h_{2}& =2(\omega _{0}^{2}L)^{-1}\cos k-2(\omega _{0}^{2}L)^{-1}-(\omega
_{0}^{2}L_{0})^{-1},  \label{TriL}
\end{align}%
instead of Eqs.(\ref{TopoL-f}). All other equations are unmodified except
that the chemical potential is given by%
\begin{equation}
\mu =-2C-C_{0}.  \label{EqAA}
\end{equation}%
The system is in the trivial phase since $\left\vert \mu \right\vert
>\left\vert 2t\right\vert $ is satisfied.

\subsection{TCU (topology-control unit)}

\label{TCU}

We have shown that the topological (trivial) segment is realized in the
right-hand (left-hand) side of Fig.\ref{FigCircuitTT}(a). These two segments
are switched from one to another by interchanging inductors $L_{0}$ and
capacitors $C_{0}$. It is remarkable that we can make a portion of the chain
topological or trivial simply by the interchange of $L_{0}$ and $C_{0}$. We
have introduced the symbol of TCU to represent this operation.

\begin{figure}[t]
\centerline{\includegraphics[width=0.48\textwidth]{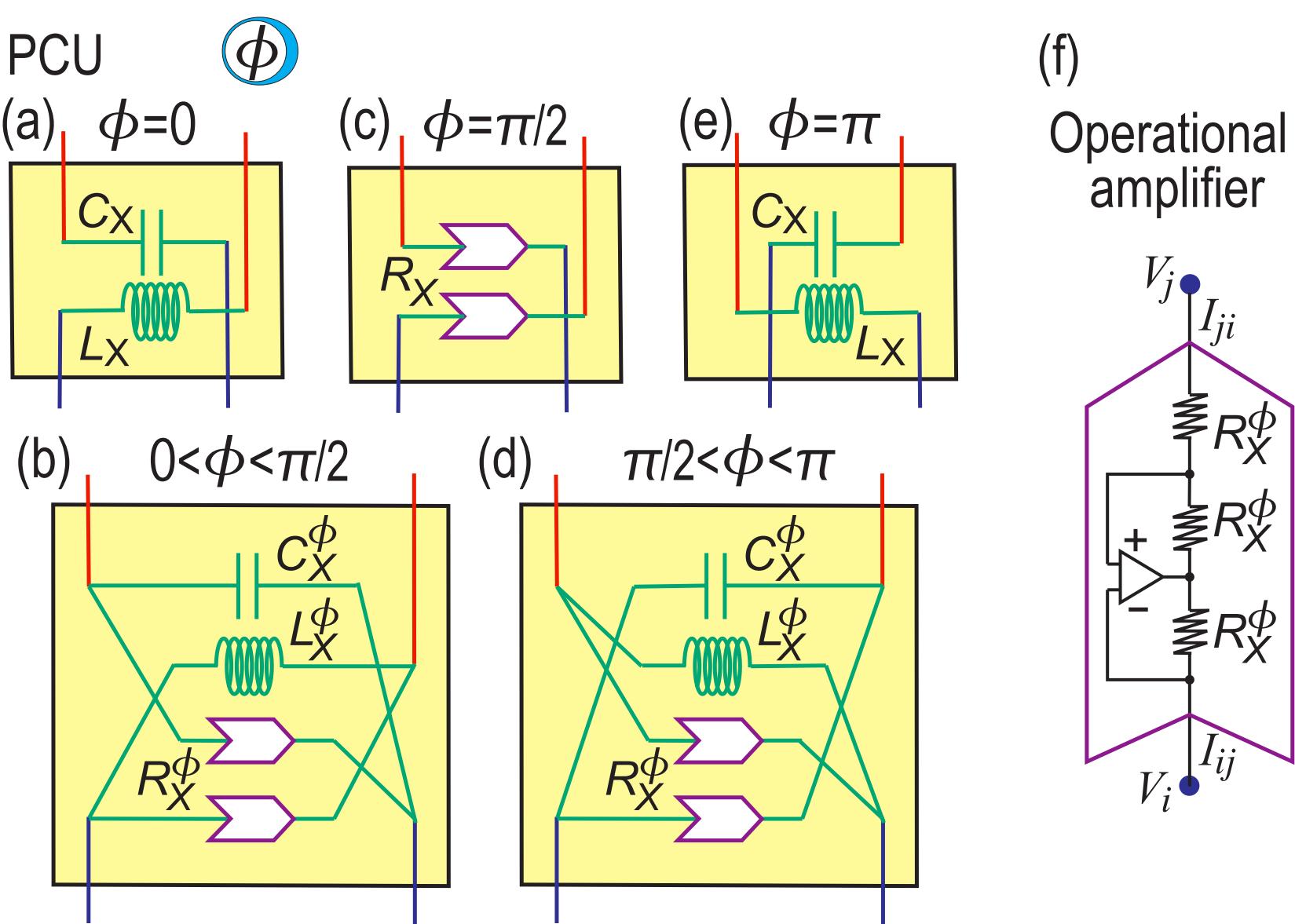}}
\caption{Illustrations of a PCU. (a)$\sim $(e) Structure of a PCU for
various values of $\protect\phi $. It consists of capacitance $C_{X}^{%
\protect\phi }$, inductance $L_{X}^{\protect\phi }$ and operational
amplifiers $R_{X}^{\protect\phi }$. Their values are given by Eq.(\protect
\ref{RCL}) to produce the "superconducting" phase $\protect\phi $. (f)
Structure of operational amplifier\protect\cite{Hofmann}.}
\label{FigPhase}
\end{figure}

\subsection{PCU (phase-control unit)}

\label{PCU}

As we have stated, Fig.\ref{FigCircuitTT}(a) is for the Kitaev $H_{\text{K}%
}^{y}$ model (\ref{Hy}) by setting $\phi =0$ in the Kitaev $H_{\text{K}}$
model (\ref{BdG}). The phase choice $\phi =0$ is made by the setting of the
pairing interactions between the two main channels shown in green in Fig.\ref%
{FigCircuitTT}(b). We have made this point explicit in Fig.\ref{FigCircuitTT}%
(c) and (e), which is equivalent to Fig.\ref{FigCircuitTT}(b), by
introducing the symbol of PCU at $\phi =0$. It is composed of the capacitor $%
C_{X}$ and the inductor $L_{X}$.

It is necessary to include the phase degree of freedom associated with $\phi 
$ to make a braiding. We have considered the case with $\phi =\pi /2$ in a
previous work\cite{EzawaMajo}, where we have used operational amplifies $%
R_{X}$. An operational amplifier is illustrated in Fig.\ref{FigPhase}(f),
which acts as a negative impedance converter with current inversion\cite%
{Hofmann}. In the operational amplifier, the resistance depends on the
current flowing direction; $R_{X}$ for the forward flow and $-R_{X}$ for the
backward flow with the convention that $R_{X}>0$.

We illustrate PCU at $\phi =0$, $0<\phi <\pi /2$, $\pi /2$, $\pi /2<\phi
<\pi $ and $\phi =\pi $ in Fig.\ref{FigPhase}(a)$\sim $(e). We explain how
the circuit Laplacian and the electric circuit are modified for each case.
The structure of PCU is determined only by modifying the pairing
interactions between the two main channels. Hence, the diagonal components $%
h_{1}$ and $h_{2}$ are not affected in the circuit Laplacian (\ref{EqC}).

(i) At $\phi =\pi $, PCU is shown in Fig.\ref{FigPhase}(a). The capacitors $%
C_{X}$ and the inductors $L_{X}$ are interchanged as compared with that at $%
\phi =0$. The circuit Laplacian is given by replacing Eqs.(\ref{TopoL-g})
with%
\begin{align}
g_{1}& =(\omega _{0}^{2}L_{X})^{-1}e^{ik}-C_{X}e^{-ik},  \notag \\
g_{2}& =-C_{X}e^{ik}+(\omega _{0}^{2}L_{X})^{-1}e^{-ik}.
\end{align}

(ii) At $\phi =\pi /2$, PCU is shown in Fig.\ref{FigPhase}(c). The circuit
is constructed with the use of operational amplifiers only, and the circuit
Laplacian is given by%
\begin{equation}
g_{1}=g_{2}=2(\omega _{0}R_{X})^{-1}\sin k.  \label{pi2g1}
\end{equation}

(iii) For $0\leq \phi \leq \pi /2$, PCU is shown in Fig.\ref{FigPhase}(b).
It is necessary to use $C_{X}$, $L_{X}$ and $R_{X}$ as a function of $\phi $
to generate the Kitaev model with $\phi $. We study the case of the
topological phase explicitly. The circuit Laplacian is given by replacing
Eqs.(\ref{TopoL-g}) with 
\begin{align}
g_{1}& =-C_{X}^{\phi }e^{ik}+(\omega _{0}^{2}L_{X}^{\phi
})^{-1}e^{-ik}+2(\omega _{0}R_{X}^{\phi })^{-1}\sin k,  \notag \\
g_{2}& =(\omega _{0}^{2}L_{X}^{\phi })^{-1}e^{ik}-C_{X}^{\phi
}e^{-ik}+2(\omega _{0}R_{X}^{\phi })^{-1}\sin k.  \label{pi2g}
\end{align}%
By requiring Eqs.(\ref{OmegaLC}) and%
\begin{equation}
C_{X}^{\phi }=C_{X}\left\vert \cos \phi \right\vert ,\quad L_{X}^{\phi }=%
\frac{L_{X}}{\left\vert \cos \phi \right\vert },\quad R_{X}^{\phi }=\frac{%
R_{X}}{\left\vert \sin \phi \right\vert },  \label{RCL}
\end{equation}%
the circuit Laplacian is reduced to%
\begin{equation}
J=\left[ 2C(1-\cos k)-C_{0}\right] \sigma _{z}+2\left( \frac{\sqrt{LC}}{%
R_{X}^{\phi }}\sigma _{x}+C_{X}^{\phi }\sigma _{y}\right) \sin k.
\end{equation}%
Formulas (\ref{RCL}) are valid for arbitrary $\phi $. In particular, when we
set $\phi \rightarrow 0$, all these equations are reduced to those in Sec.%
\ref{TCU}. On the other hand, when we set $\phi \rightarrow \pi /2$, Eqs.(%
\ref{pi2g}) are reduced to Eqs.(\ref{pi2g1}). A similar analysis is made
with respect to the trivial phase.

(iv) For $\pi /2\leq \phi \leq \pi $, PCU is shown in Fig.\ref{FigPhase}(d).
We study the topological phase explicitly. The circuit Laplacian is given by
replacing Eqs.(\ref{TopoL-g}) with%
\begin{align}
g_{1}& =(\omega _{0}^{2}L_{X}^{\phi })^{-1}e^{ik}-C_{X}^{\phi
}e^{-ik}+2(\omega _{0}R_{X}^{\phi })^{-1}\sin k,  \notag \\
g_{2}& =-C_{X}^{\phi }e^{ik}+(\omega _{0}^{2}L_{X}^{\phi
})^{-1}e^{-ik}+2(\omega _{0}R_{X}^{\phi })^{-1}\sin k.  \label{pi2g2}
\end{align}%
The circuit Laplacian is reduced to%
\begin{equation}
J=\left[ 2C(1-\cos k)-C_{0}\right] \sigma _{z}+2\left( \frac{\sqrt{LC}}{%
R_{X}^{\phi }}\sigma _{x}-C_{X}^{\phi }\sigma _{y}\right) \sin k,
\end{equation}%
by requiring Eqs.(\ref{RCL}). We note that, when we set $\phi \rightarrow
\pi /2$, Eqs.(\ref{pi2g2}) are reduced to Eqs.(\ref{pi2g1}).

Consequently, when we use $C_{X}^{\phi }$, $L_{X}^{\phi }$ and $R_{X}^{\phi
} $ defined by Eqs.(\ref{RCL}) within PCU, the Kitaev model with arbitrary $%
\phi $ is realized in electric circuits.

\begin{figure}[t]
\centerline{\includegraphics[width=0.48\textwidth]{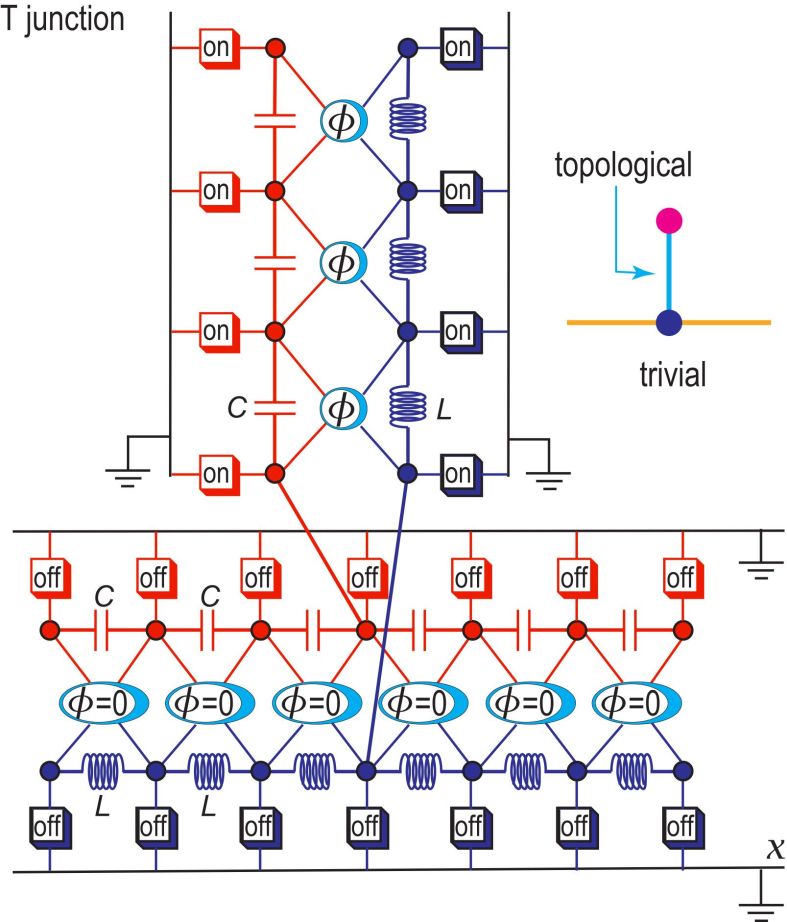}}
\caption{(a) Illustrations of an electric circuit for a T-junction in terms
of TCUs and PCUs. The horizontal part is essentially the same as the one in
Fig.\protect\ref{FigCircuitTT} except that it is entirely in the trivial
phase. This circuit is for the configurations in Fig.\protect\ref{FigTBraid}%
(d) and (e), where the horizontal leg-1 and leg-2 are trivial while the
vertical leg-3 is topological. }
\label{FigTCircuit}
\end{figure}

\subsection{T-junction}

A T-junction may be designed in electric circuits as in Fig.\ref{FigTCircuit}%
. This circuit is for the configurations in Fig.\ref{FigTBraid}(d), where
the horizontal leg-1 and leg-2 are trivial while the vertical leg-3 is
topological. When all TCUs are set on, it is for the configurations in Fig.%
\ref{Fig4Braid}(d), where all three legs are topological.

During a braiding process, it is necessary to control $\phi $ continuously
from $\phi =0$ to $\phi =\pi $ to proceed from (d) to (e) in Fig.\ref%
{FigTBraid} and Fig.\ref{Fig4Braid}. A similar control is necessary from (h)
to (i) in these figures. Such an operation is made possible by using a
rotary switch tuning variable parameters $R^{\phi }$, $C_{X}^{\phi }$ and $%
L_{X}^{\phi }$ within a PCU according to the formula (\ref{RCL}), as we
discuss later.

\begin{figure}[t]
\centerline{\includegraphics[width=0.48\textwidth]{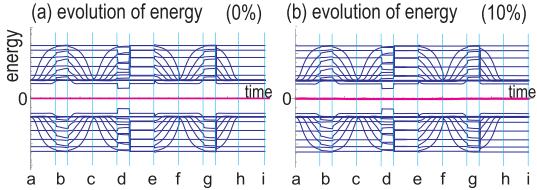}}
\caption{Evolution of the energy as a function of time, (a) without any
randomness (0\%), and (b) with 10\% randomness on the on-site potential.
Figure (a) is the same as Fig.\protect\ref{FigXBraid}(b). The energy of the
edge states (in red) are found to slightly deviate from zero by randomness
in (b). }
\label{FigPHS}
\end{figure}

\begin{figure}[t]
\centerline{\includegraphics[width=0.48\textwidth]{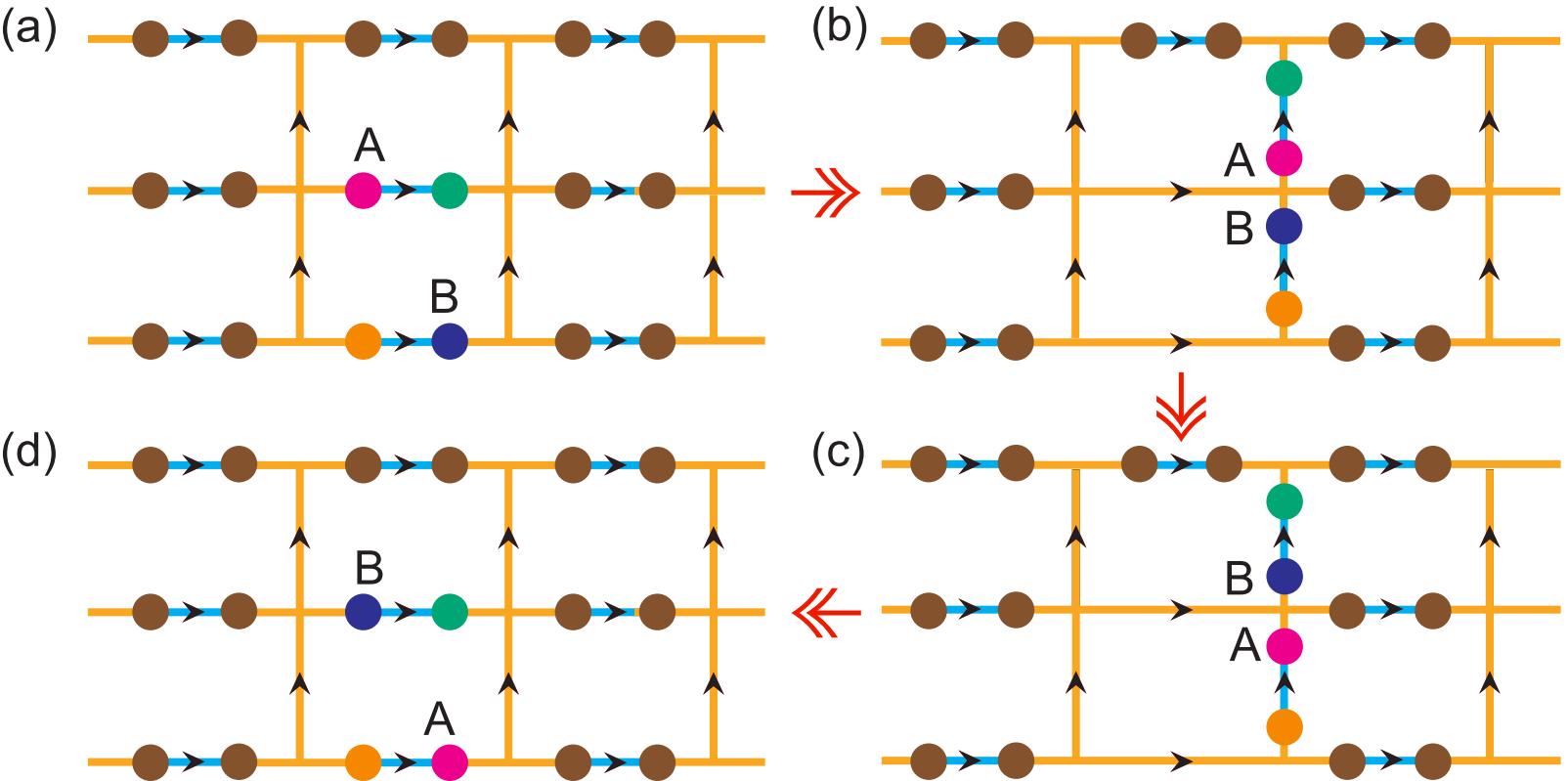}}
\caption{Non-Abelian braiding of Majorana states deposited on a square
network. Topological segments (cyan) are created on horizontal parallel
Kitaev chains, which are connected by vertical parallel Kitaev chains. A
crossroads may be used as a T-junction for two edges to braid. In this
example two edges A (in red) and B (in blue) on different parallel chains
are braided by following the steps (a), (b), (c) and (d). }
\label{FigSquareBraid}
\end{figure}

\section{Discussions}

We have explored the physics associated with topological edge states in an
electric circuit whose circuit Laplacian is equivalent to the Hamiltonian
for the Kitaev $p$-wave superconductor. The circuit contains two main
channels (capacitor channel and inductor channel) corresponding to the
electron band and hole band. The position and the phase of an edge soliton
are externally controlled by TCUs and PCUs, respectively, and observable by
an impedance peak.

The braiding of edge states is performed with the aid of T-junction. In
particular, we have derived the one-qubit (two-qubit) gate resulting from
the braiding of two edge states across a topological (trivial) segment. The
results agree precisely with those obtained based on Majorana fermions in
superconducting systems. Consequently, quantum gates based on our electric
circuits will be entirely equivalent to the standard quantum gates based on
Majorana fermions. A merit of electric circuit realization is that we can
exactly set the critical point $t=\Delta $, which is practically impossible
in topological superconductors. Recall that the edge states are exactly
localized when $t=\Delta $.

We have shown that quantum gates are constructed in electric circuits
provided PHS is intact. In actual electric circuits, PHS will be broken
weakly due to the randomness, which acts as the on-site potential
randomness. We show the energy spectrum evolution in the presence of the
10\% on-site potential randomness in Fig.\ref{FigPHS}. Although the edge
states acquire slight non-zero values, they are well separated from the bulk
spectrum. They evolve smoothly as a function of time since the randomness is
solely determined by sample elements which are fixed during time evolution.
Furthermore, it is possible to make fine tuning of sample elements in the
electric circuit for a practical degeneracy of edge states. We should
mention that, once fine tuning is made, we can use it for good.

By generalizing a one-dimensional array of the T-junctions to the two
dimensions, it becomes possible to braid edge states on a square lattice,
which is illustrated in Fig.\ref{FigSquareBraid}. In the same way, we can
generalize them to the three dimensions, where edge states are deposited on
a cube.

Some ingenuity might be necessary in actual implementation of integrated
circuits. Varactor (variable capacitance diode) will be useful to control
the capacitance\cite{Garcia}. Inductors may be displaced by simulated
inductors\cite{SimL} with the use of operational amplifiers.

We have pursued the parallel between the Kitaev superconductor model hosting
Majorana fermions and the Kitaev electric-circuit chain governed by the
Kirchhoff laws. As we have shown, it is possible to introduce Majorana
operators in the electric circuit. However, this does not mean that it is
possible to construct a whole quantum system. A quantum system has two main
properties: (i) A linear algebra structure and (ii) the contraction of the
wave function. The electric-circuit system has the property (i) but not the
property (ii). Let us explain a bit more in details.

Electric circuits can simulate qubits, unitary transformation, superposition
and entanglement, as we have shown. This is so because they require only a
linear algebra structure, which exists also in the Kirchhoff laws. On the
other hand, they cannot simulate the contraction of the wave function,
because there is no probabilistic phenomena in electric circuits.
Accordingly, we cannot simulate such quantum algorithms that use the
contraction of the wave function. In addition, electric circuits cannot
simulate quantum communications such as quantum teleportation by the same
reason. However, it is not so serious as a topological quantum computer.
Indeed, most quantum algorithms based on the braiding of Majorana fermions
do not require the contraction of the wave function.

The author is thankful to E. Saito, Y. Mita and N. Nagaosa for helpful
discussions on the subject. This work is supported by the Grants-in-Aid for
Scientific Research from MEXT KAKENHI (Grants No. JP17K05490, No. JP15H05854
and No. JP18H03676). This work is also supported by CREST, JST (JPMJCR16F1
and JPMJCR1874).

\end{document}